%% file: jcapexample.tex
\documentclass[a4paper,11pt]{article}
\pdfoutput=1 % if your are submitting a pdflatex (i.e. if you have
\usepackage{jcappub} % for details on the use of the package, please
\usepackage[T1]{fontenc} % if needed
\usepackage[normalem]{ulem}
\usepackage{multirow}
\usepackage{hyperref}
\usepackage{doi} % for doi hyperlink
\usepackage{subcaption}
\usepackage{verbatim}
\usepackage{soul}
\usepackage{graphics}
\usepackage{hhline}
\usepackage{cleveref}
\usepackage{graphicx}
\usepackage{xcolor}
\newcommand*{\bo}{\boldsymbol}
\newcommand*{\mr}{\mathrm}
\newcommand*{\hmpc}{h \, \mr{Mpc}^{-1}}
\newcommand*{\kmax}{k_\mr{max}}

\title{\boldmath Mitigation of nonlinear galaxy bias with a theoretical-error likelihood}

\author[a,1]{A. Aires,\note{Corresponding author.}}
\author[b,c]{N. Kokron,}
\author[a,d]{R. Rosenfeld,}
\author[e,f]{F. Andrade-Oliveira,}
\author[g]{V. Miranda}

\affiliation[a]{Institute for Theoretical Physics, State University of Sao Paulo, 271 Dr. Bento Teobaldo Ferraz, Sao Paulo, Brazil}

\affiliation[b]{School of Natural Sciences, Institute for Advanced Study, 1 Einstein Drive, Princeton, NJ, 08540, USA}
\affiliation[c]{Department of Astrophysical Sciences, Peyton Hall, Princeton University, 4 Ivy Lane, Princeton, NJ 08544, USA}

\affiliation[d]{ICTP South American Institute for Fundamental Research, 271 Dr. Bento Teobaldo Ferraz, Sao Paulo, Brazil}
\affiliation[e]{Department of Physics, University of Michigan, Ann Arbor, MI 48109, USA}
\affiliation[f]{Physik-Institut — University of Zurich, Winterthurerstrasse 190, 8057 Zurich, Switzerland}
\affiliation[g]{C. N. Yang Institute for Theoretical Physics, Stony Brook University, Stony Brook, NY 11794, USA}
\emailAdd{abdias.aires@unesp.br, kokron@astro.princeton.edu, rogerio.rosenfeld@unesp.br, felipe.andradeoliveira@physik.uzh.ch, vivian.miranda@stonybrook.edu}

\abstract{Stage-IV galaxy surveys will measure correlations at small cosmological scales with high signal-to-noise ratio. One of the main challenges of extracting information from small scales is devising accurate models, as well as characterizing the theoretical uncertainties associated with any given model. In this work, we explore the mitigation of theoretical uncertainty due to nonlinear galaxy bias in the context of photometric 2$\times$2-point analyses. We consider linear galaxy bias as the fiducial model and derive the contribution to the covariance matrix induced by neglected higher-order bias. We construct a covariance matrix for the theoretical error in galaxy clustering and galaxy-galaxy lensing using simulation-based relations that connect higher-order parameters to linear bias. To test this mitigation model, we apply the modified likelihood to 2$\times$2-point analyses based on two sets of mock data vectors: (1) simulated data vectors, constructed from those same relations between bias parameters, and (2) data vectors based on the {\tt AbacusSummit} simulation suite. We then compare the performance of the theoretical-error approach to the commonly employed scale cuts methodology. We find most theoretical-error configurations yield results equivalent to the scale cuts in terms of precision and accuracy, and in some cases, especially with the first data set, they produce significantly stronger bounds on cosmological parameters. These results are independent of the maximum scale $\kmax$ in the analysis with theoretical error. We notice the relative performance of the theoretical-error approach depends mostly on the choice of the covariance-matrix diagonal. The scenarios where linear bias supplemented by theoretical error is unable to recover unbiased cosmology, which are mainly observed with the second data set, are connected to inadequate modeling of the $gg$-$g\kappa$ covariance of theoretical error. This form of cross-probe covariance has not been considered in previous works. We additionally highlight a sensitivity of the construction to off-diagonal correlations of theoretical error. In view of its removing the ambiguity in the choice of $\kmax$, as well as the possibility of attaining higher precision than the usual scale cuts, we consider this method to be promising for analyses of LSS in upcoming photometric galaxy surveys.}

\keywords{cosmological parameters from LSS, galaxy surveys, cosmological simulations, power spectrum}

\begin{document}
\maketitle
\flushbottom

\input{intro}
\input{method}

\input{TE_models}

\input{lik_analysis}

\input{results}

\input{conclusion}

\vspace{5mm}
\section*{Acknowledgements}

This research was supported by resources supplied by the Center for Scientific Computing (NCC/GridUNESP) of the São Paulo State University (UNESP). The work of AA was financed by the Coordenação de Aperfeiçoamento de Pessoal de Nível Superior - Brasil (CAPES) - Finance Code 001 and the Brazilian Agency CNPq. NK acknowledges support from NSF award AST-2108126 and from the Fund for Natural Sciences of the Institute for Advanced Study. NK and FAO would like to thank the International Center for Theoretical Physics – South American Institute for Fundamental Research (ICTP-SAIFR) for their hospitality during part of the completion of this work. ICTP-SAIFR is funded by FAPESP grant 2021/14335-0. The work of RR is supported in part by the São Paulo Research Foundation (FAPESP) through grant 2021/10290-2 and by the Brazilian Agency CNPq through a productivity grant 311627/2021-8 and the INCT of the e-Universe. FAO acknowledges the support from the Swiss National Science Foundation (SNSF) through grant No. 231766. 
\appendix
\section{A toy model for the running of SNR with $\Delta \ell$}
\label{sec:appendix}
We consider a theoretical error problem with only two angular multipoles, $\ell_1$ and $\ell_2$. We define the Gaussian and TE covariance, respectively, as 
\begin{align}
{\rm Cov}^G = 
    \begin{pmatrix}
        \sigma_{\ell_1}^2 & 0 \\ 
        0 & \sigma_{\ell_2}^2 
    \end{pmatrix},
    \quad 
    {\rm Cov}^{\rm TE} = 
    \begin{pmatrix}
        E_{\ell_1}^2 & \rho E_{\ell_1} E_{\ell_2} \\
        \rho E_{\ell_1} E_{\ell_2} & E_{\ell_2}^2
    \end{pmatrix},
\end{align}
where $\sigma_\ell$ is the uncertainty due to cosmic variance, $E_\ell$ is the envelope, and $\rho$ is the TE correlation coefficient, which we normally write as $\rho = E_{\ell_1} E_{\ell_2} e^{(\ell_1 - \ell_2)^2/2\Delta \ell^2}$. We compute the signal-to-noise ratio of the Gaussian+TE covariance as the quadratic form 
\begin{equation}
    {\rm SNR}(\rho) = (C_{\ell_1}, C_{\ell_2})^T \cdot \left [{\rm Cov}^G + {\rm Cov}^{\rm TE} \right ]^{-1} \cdot (C_{\ell_1}, C_{\ell_2}),
\end{equation}
where we have made the dependence on $\rho$ (or, equivalently, $\Delta \ell$) explicit. If we fix the envelope, $\rho$ is the only free parameter remaining to be fixed in the theoretical-error model. The SNR for this problem has, formally, two extrema. Searching for them,
\begin{equation}
    \left. \frac{\partial {\rm SNR}(\rho)}{\partial \rho} \right |_{\rho_*} = 0,
\end{equation}
we find
\begin{equation}
    \rho_{*,1} = \frac{C_{\ell_2}}{E_{\ell_2}} \frac{\sigma_{\ell_1}}{C_{\ell_1}} \left ( \frac{E_{\ell_1}}{\sigma_{\ell_1}} + \frac{\sigma_{\ell_1}}{E_{\ell_1}}\right ), \quad \rho_{*,2} = \frac{C_{\ell_1}}{E_{\ell_1}} \frac{\sigma_{\ell_2}}{C_{\ell_2}} \left ( \frac{E_{\ell_2}}{\sigma_{\ell_2}} + \frac{\sigma_{\ell_2}}{E_{\ell_2}}\right ).
\end{equation}

By computing the eigenvalues of $\mr{Cov^G + Cov^{TE}}$ as functions of $\rho$, we can show that either $\rho_{*,1}$ or $\rho_{*,2}$ (but not both) is in the range of $\rho$ for which the matrix is positive-definite. That is, the function SNR($\rho$) has only one extremum in the allowed range of $\rho$. We can also check this extremum is a minimum. Hence, for $\Delta \ell$ values such that the resulting $\rho$ satisfies $\rho \geq {\rm Min}(\rho_{*,1}, \rho_{*,2})$, the SNR \emph{increases}  as we correlate $\ell$-bins more strongly. That is, the theoretical-error procedure correlates variables so highly that they begin to \emph{inform} each other, and the overall signal-to-noise increases. Writing $\rho$ in its squared-exponential form, we can explicitly relate a $\Delta\ell$ to $\rho_{*}$. Broadly speaking, the presence of this minimum in the SNR with theoretical error explains the shape of figure \ref{fig:snr_vs_deltak}.

Some intuition on the exact value of $\rho_*$ can be gleaned from parameterizing the off-diagonal contribution in terms of the Gaussian uncertainty: 
\begin{equation}
    \rho E_{\ell_1} E_{\ell_2} = \tilde{\rho} \sigma_{\ell_1}\sigma_{\ell_2}.
\end{equation}
In this case the condition for $\tilde{\rho}_{*,1}$ which extremizes SNR reads 
\begin{equation}
\label{eq:rhotilde_gausscor}
    \tilde{\rho}_{*,1} = \frac{C_{\ell_2}}{\sigma_{\ell_2}}\frac{\sigma_{\ell_1}}{C_{\ell_1}} \left ( 1 + \frac{E_{\ell_1}^2}{\sigma_{\ell_1}^2} \right ). 
\end{equation}

That is, when the correlation coefficient exceeds the ratio of SNRs of the Gaussian multipoles, the correlation becomes informative. This equation also highlights that $\tilde{\rho}_*$ increases with the envelope. A larger envelope accommodates a larger correlation length in the off-diagonal before the correlation becomes informative.

\section{Projection effects}
\label{sec:app_proj}

\begin{figure}[tbp]
\centering % \begin{center}/\end{center} takes some additional vertical space
\includegraphics[width=0.5\textwidth,clip]{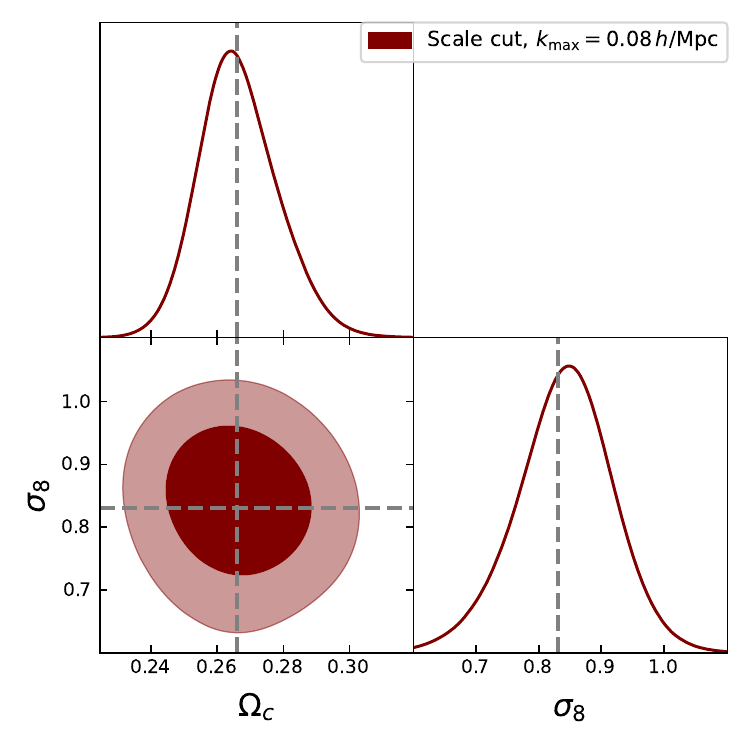}
\hfill
\caption{\label{fig:sdv_proj} Analysis with a scale cut at $\kmax = 0.08 \hmpc$ and a data vector without nonlinear bias. The inference with a linear bias model attains a good fit with the data and exhibits very low tension (0.03$\sigma$) in the marginalized 2D posterior of $\sigma_8$ and $\Omega_c$.}
\end{figure}

Our comparison of TE and scale cut is made in terms of tension and precision in the marginalized 2D posterior of $\sigma_8$ and $\Omega_c$. This raises the issue of projection effects. To check that they are small, we run an MCMC analysis with a scale cut at $\kmax = 0.08 \hmpc$ and a data vector without nonlinear bias. The $b_1$ values are the same as described in subsection \ref{sec:lik_analysis_sdv} for simulated data vectors, namely, 0.7 and 1.0 for bins 0 and 1, respectively. As expected, an inference with the linear bias model attains a good fit with the data and exhibits very low tension (0.03$\sigma$) in the marginalized 2D posterior. Projection effects are small. This is consistent with the geometry of our parameter space, which is smaller and simpler than the one adopted, for example, in the DES Y3 analysis \cite{krause2021dark}. The marginalized posterior is presented in figure \ref{fig:sdv_proj}.

\section{Noisy data vectors}
\label{sec:app_noise}

To test robustness with respect to noise, we sample noisy data vectors from a Gaussian distribution with the ``typical bias'' vector as mean (subsection \ref{sec:lik_analysis_sdv}). We use the Gaussian covariance matrix \eqref{eq:cov_d} without marginalization 
over theoretical errors, containing $gg$--$gg$, $g\kappa$--$g\kappa$, and $gg$--$g\kappa$ correlations. The TE configuration has
a ``95\%'' envelope, $\Delta k = 0.05 \hmpc$, zero $gg$--$g\kappa$ TE covariance, and $\kmax = 1\hmpc$. The scale cut has $\kmax = 0.08 \hmpc$.

\begin{figure}
\centering
\begin{subfigure}{0.50\textwidth}
  \centering
  \includegraphics[width=\linewidth]{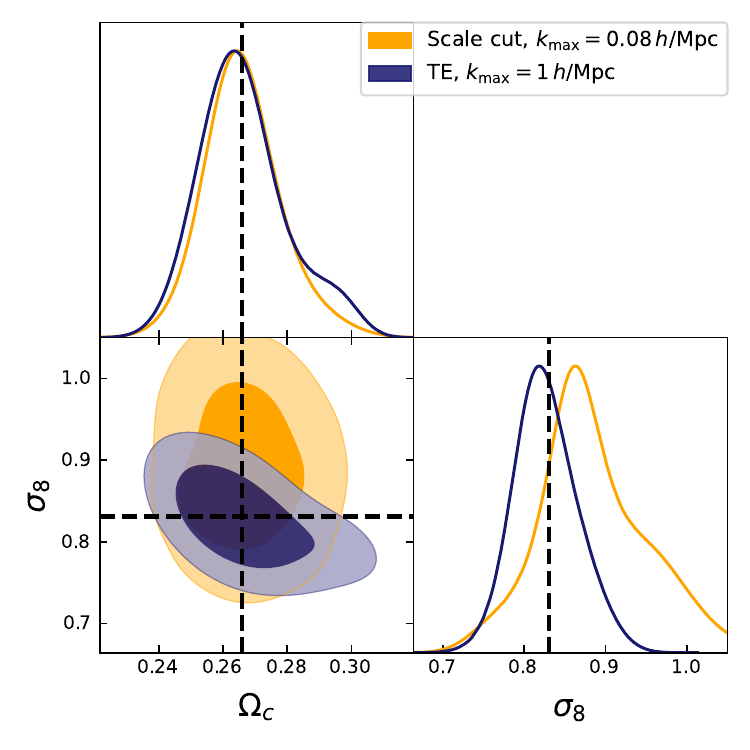}
  \caption{\label{fig:sdv_noise1}``Noise 3''}
\end{subfigure}%
\begin{subfigure}{0.50\textwidth}
  \centering
  \includegraphics[width=\linewidth]{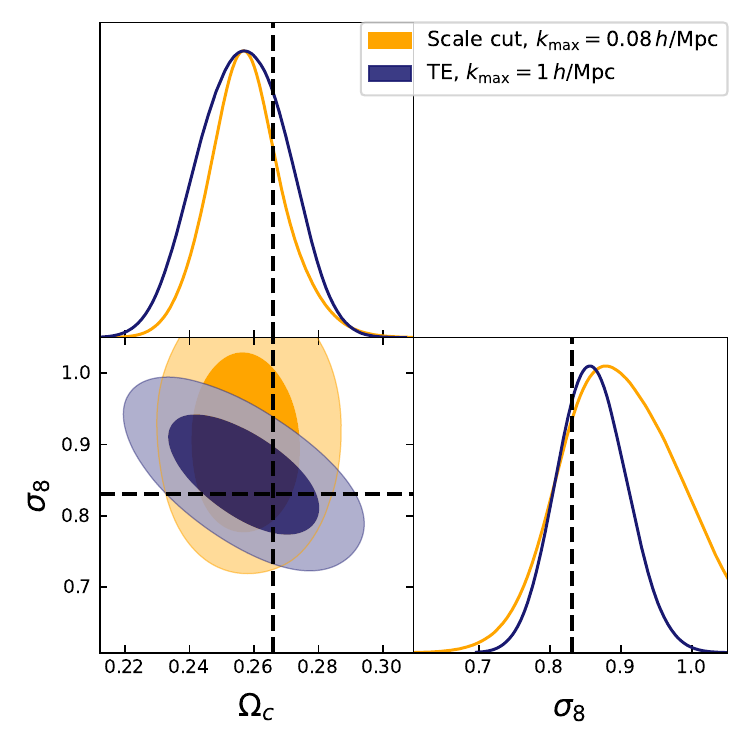}
  \caption{\label{fig:sdv_noise2}``Noise 4''}
\end{subfigure}
\caption{\label{fig:sdv_noise} Contour plots of TE and scale cut corresponding to two
noisy realizations of the ``typical bias'' data vector. The TE configuration is specified
by a ``95\%'' envelope, $\Delta k = 0.05\hmpc$, zero $gg$--$g\kappa$ TE covariance, and 
$\kmax = 1\hmpc$. The scale cut has $\kmax = 0.08 \hmpc$. TE produces a better fit with 
the data vector, as 
measured by a lower reduced $\chi^2$. It achieves stronger constraining power than
the scale cut for similar or higher accuracy. }
\end{figure}

The results are presented in table \ref{tab:noise}. TE exhibits greater robustness to noise. It attains similar or higher 
accuracy than the scale cut with greater constraining power, as expressed in its larger FoM. The reduced $\chi^2$
remains below or equal to 0.70 for all tested vectors, whereas for the scale cut it exceeds 1.40 in three of the five cases. 
Posterior plots are displayed in figure \ref{fig:sdv_noise}.

\input{noise_table}

\bibliographystyle{JHEP}
\bibliography{jcapexample} % Entries are in the refs.bib file

\end{document}

%% file: intro.tex
\section{Introduction}
\label{sec:intro}

The precision level of the cosmological information from large galaxy surveys has become comparable to that obtained from the cosmic microwave background (CMB). Theoretical analyses using data from these surveys can be more complex than those that rely on the CMB.
This is due to several non-trivial effects, such as nonlinearities in the perturbations, effects of baryonic feedback on the power spectrum, galaxy bias, and the intrinsic alignment of galaxies \cite{krause2021dark, Moreira_2021, baldauf2016lss, Chen_2020}. They are usually incorporated in the analyses either (i) by making a judicious choice of scales where they are under control or (ii) by the addition of extra phenomenological parameters to describe them \cite{krause2021dark}. The former approach requires a careful study of what scales can be safely kept in the analyses, with a proper definition of what is meant by ``safe''. It involves disregarding a fraction of the data that does not pass this requirement, in what is known as the ``scale cut'' procedure. The second approach often has the disadvantage of introducing a large number of nuisance parameters, which entails the computational cost of sampling over a high-dimensional parameter space. For example, a naive extension of linear galaxy bias to one-loop non-linear bias for a photometric survey introduces on the order of 4 to 5 additional parameters \emph{per redshift bin} \cite{Desjacques:2016bnm, LSSTDarkEnergyScience:2023qfp}. Beyond the challenges of sampling such a large parameter space, improperly assigned priors can also aggravate volume projection effects~\cite{Lemos_2021, Simon_2023, guachalla2024informedtotalerrorminimizingpriorsinterpretable, raveri2024understandingposteriorprojectioneffects}. 

An alternative approach is sometimes referred to as the theoretical-error method (TE) \cite{baldauf2016lss, Chudaykin_2021, Moreira_2021}. The main idea is to statistically estimate the errors incurred from neglecting a more sophisticated model and incorporate them in an augmented covariance matrix. This matrix smoothly reduces the signal-to-noise contribution from scales where the simple model is not sufficiently accurate. This approach has been recently studied in the mitigation of baryonic feedback effects in galaxy shear analyses \cite{Moreira_2021, maraio2024mitigatingbaryonfeedbackbias}, as well as in the modeling of nonlinear dark matter and galaxy power spectra in real and redshift spaces in spectroscopic surveys \cite{Chudaykin_2021}. A similar methodology has been applied to the error in the inference of cosmological parameters through neural networks \cite{Grandon2022Bayesian}. In this work we develop and test the TE method in the case of galaxy bias, the relation between the dark matter and galaxy distributions. Specifically, we apply it to the analyses of galaxy-galaxy and galaxy-galaxy lensing 2-point correlation functions, the so-called 2$\times$2-point analyses, consistent with photometric galaxy surveys such as the Rubin Observatory's Legacy Survey of Space and Time (LSST). The pure galaxy-lensing angular correlation function, sometimes called $\kappa \kappa$, is not sensitive to galaxy bias and therefore it is not considered in this work.

Several prescriptions have been proposed to model galaxy bias. As reviewed in \cite{Desjacques:2016bnm}, they can be broadly divided into phenomenological and perturbative models. The former relies on heuristic prescriptions to populate dark matter halos with galaxies drawn from a halo occupation distribution (HOD). Perturbative models, on which we focus in this work, use perturbative expansions packaged in an effective field theory (EFT) framework to address galaxy bias. The performance of different perturbative galaxy bias models in the context of the LSST has been recently presented in \cite{LSSTDarkEnergyScience:2023qfp}.

Specifically, we consider the theoretical uncertainty arising from restricting a 2$\times$2-point analysis to one (linear) bias parameter. To model the augmented covariance matrix, we suppose the true data vectors to be described by a second-order Lagrangian EFT bias expansion, supported by a one-loop matter power spectrum in Lagrangian perturbation theory. We estimate the range of possible theoretical errors by varying the second-order bias parameters according to simulation-based relations. These capture the physical correlations between linear and higher-order bias allowed by models of galaxy formation.

We test this approach by applying it to mock cosmological analyses in a simplified setting. These are based on two sets of mock data vectors: (i) simulated
data vectors, generated from the second-order bias model itself, and (ii) vectors measured from galaxy samples based on the {\tt AbacusSummit} N-body simulation suite \cite{Garrison_2018} and constructed in the DESC Bias Challenge \cite{LSSTDarkEnergyScience:2023qfp}. In both cases, we compare the results of statistical inference in the TE approach to those obtained with scale cuts.

This paper is organized as follows. In section \ref{sec:method}, we review a modified likelihood function that takes into account theoretical uncertainty. We also provide a brief review of the perturbative bias expansion and introduce the 2$\times$2-point data vectors, as well as the Gaussian covariance matrices used in this work. In section \ref{sec:envelopes}, we specify the models of Gaussian probability distributions of theoretical error. We introduce the test methodology in section \ref{sec:lik_analysis} and present our results in section \ref{sec:results}. Section \ref{sec:conc} contains our conclusions.

%% file: method.tex
\section{Methodology}

In this section we discuss several aspects of methodology. We begin by constructing a modified likelihood function that takes account of theoretical uncertainty, as introduced in \cite{baldauf2016lss}. We then specify the Lagrangian bias expansion to be used in this work and define 2$\times$2-point data vectors.

\label{sec:method}

\subsection{Marginalization over theoretical error}
\label{sec:theo_review}

Given an inference problem with data vector $\bo d$ and theoretical vector $\bo t$, we treat the theoretical error, $\bo e \equiv \bo d - \bo t$, as a random variable with an yet-unspecified probability distribution $P(\boldsymbol{e})$ \cite{baldauf2016lss}. The Gaussian likelihood of the data given the theory and the theoretical error is:

\begin{equation}
    \mathcal{L}(\boldsymbol{d} \, | \, \boldsymbol{t},\boldsymbol{e}) = \frac{1}{\sqrt{(2\pi)^m |\mr{Cov}_d|}} \; \mathrm{exp} \left[-\frac{1}{2} (\boldsymbol{d}-\boldsymbol{t}-\boldsymbol{e}) \mr{Cov}_d^{-1} (\boldsymbol{d}-\boldsymbol{t}-\boldsymbol{e})^\mathrm{T}) \right],
\end{equation}

\noindent where $\mr{Cov}_d$ is the covariance matrix and $m$ is the dimension of $\bo d$, $\bo t$, and $\bo e$. As a further step, we obtain the likelihood of the data given the theory, irrespective of the particular value of theoretical error, by marginalization over the latter:

\begin{equation}
    \mathcal{L}(\boldsymbol{d}|\boldsymbol{t}) = \int \mr d\boldsymbol e \, P(\boldsymbol{d}\,|\,\boldsymbol{t},\boldsymbol{e}) P(\boldsymbol{e}) .
\end{equation}

If we assume the error to be described by a Gaussian distribution with covariance $\mr{Cov}_e$ and mean $\boldsymbol{\bar{e}}$, the integral above yields \cite{baldauf2016lss}:

\begin{equation}
    \label{eq:mod_likelihood}
    \mathcal{L}(\boldsymbol{d}|\boldsymbol{t}) = \frac{1}{\sqrt{(2\pi)^m |\mr{Cov}_T|}} \; \mathrm{exp} \left\{ -\frac{1}{2}   \left[(\boldsymbol{d}-\boldsymbol{t}-\boldsymbol{\bar e}) \, \mr{Cov}_T^{-1} \, (\boldsymbol{d}-\boldsymbol{t}-\boldsymbol{\bar e})^\mathrm{T}\right] \right\},
\end{equation}  

\noindent with $\mr{Cov}_T = \mr{Cov}_d + \mr{Cov}_e$. The theoretical uncertainty is thus expressed in an augmented covariance matrix. 

In this work, the theoretical uncertainty arises from neglecting higher-order bias parameters in models of galaxy power spectra. Hence, we devote the next few sections to defining the variables in eq. \eqref{eq:mod_likelihood}. Specifically, the theoretical vectors $\bo t$ are linear-bias angular power spectra of galaxy clustering and galaxy-galaxy lensing. We assume the data vectors $\bo d$ to be well described by angular power spectra in the second-order bias expansion (subsection \ref{sec:lik_analysis_sdv}) or calculated from simulated galaxy samples (subsection \ref{sec:lik_analysis_abac}). The error $\bo e = \bo d - \bo t$ arises from disregarding the higher-order bias parameters in the model $\bo t$. We briefly review the bias EFT expansion in the next subsection and define the galaxy angular spectra used in this work in subsection \ref{sec:models}. We construct the augmented covariance matrix $\mr{Cov}_T$ and the mean theoretical error $\bo{\Bar e}$ in section \ref{sec:envelopes}.

\subsection{The Lagrangian bias expansion}
\label{sec:bias}

In this subsection we introduce the galaxy bias expansion to be adopted in this work. At scales larger than the typical size of a host halo, we can describe the relation between the galaxy overdensity $\delta_g$ and the matter overdensity $\delta_m$ as a local process. By the Equivalence Principle, it can only depend on the second derivatives of the gravitational potential $\partial_i\partial_j \Phi$ \cite{Desjacques:2016bnm}. Near the typical size of a host halo, however, the local approximation breaks down and the galaxy overdensity receives a contribution from the nonlocal operator $\nabla^2\delta_m$ \cite{Desjacques:2016bnm, LSSTDarkEnergyScience:2023qfp}.

Up to second order, the relevant operators in the perturbative bias expansion are $\delta_m$, $\delta_m^2$, $s^2$, and $\nabla^2\delta_m$. Here, $s^2$ is the trace squared of the tidal tensor $s^2 = s_{ij} s^{ij}$, with

\begin{equation}
    s_{ij} = \partial_i \partial_j \Phi - \delta_{ij}\nabla^2\Phi/3,
\end{equation}

\noindent whereas the matter overdensity $\delta_m$ appears in the expansion in view of its being proportional to the trace $\nabla^2\Phi$. 

In this work we adopt the Lagrangian picture. The galaxy overdensity $\delta^L_g(\bo q)$ can be established by an expansion in the initial conditions at high redshifts (Lagrangian coordinates $\bo q$), then evolved to late times by following the galaxy trajectories under gravity. The relation between the late-times ``Eulerian'' field $\delta_g(\bo x)$ and the initial Lagrangian field $\delta^L_g(\bo q)$ is:

\begin{equation}
\label{eq:lag_to_euler}
    1 + \delta_g(\bo x,z) = \int \mr d^3\bo q \, \delta_D (\bo x - \bo q - \bo \Psi(\bo q,z)) (1+\delta^L_g(\bo q)),
\end{equation}

\noindent with:

\begin{equation}
    \delta^L_g(\bo q) = b_1^L \delta_L(\bo q) + b_2^L \left[ \delta_L^2(\bo q) - \langle \delta_L^2 \rangle \right] + b^L_{s^2} \left[ s_L^2(\bo q) - \frac{2}{3}\langle \delta_L^2 \rangle \right] + b^L_{\nabla^2}\nabla^2 \delta_L(\bo q)
\end{equation}

\noindent and where $\delta_D(\bo x)$ is the Dirac delta function, $\delta_L$ is the matter overdensity in the Lagrangian frame, and $\bo \Psi$ is the Lagrangian displacement field \cite{Desjacques:2016bnm, LSSTDarkEnergyScience:2023qfp}. To make connection with statistical observables, the galaxy-galaxy power spectra $P_{gg}(\boldsymbol{k})$ and galaxy-matter power spectra $P_{gm}(\boldsymbol{k})$ are defined through:

\begin{equation}
    P_{gg}(\boldsymbol{k}) (2\pi)^3 \delta_D(\boldsymbol{k}+\boldsymbol{k}') = \langle \delta_g(\boldsymbol{k}) \delta_g(\boldsymbol{k}') \rangle,
\end{equation}

\begin{equation}
    P_{gm}(\boldsymbol{k}) (2\pi)^3 \delta_D(\boldsymbol{k}+\boldsymbol{k}') = \langle \delta_g(\boldsymbol{k}) \delta_m(\boldsymbol{k}') \rangle.
\end{equation}

By replacing eq. \eqref{eq:lag_to_euler} in the above, we find an expansion for the power spectra in terms of Lagrangian operators $\cal{O} \in $ $\{1,\delta_L,\delta^2_L,s^2_L,\nabla^2\delta_L\}$ and the corresponding bias parameters $\{b_0, b_1^L, b_2^L, b_{s^2}^L, b_{\nabla^2}^L\}$. We can write:

\begin{equation}
\label{eq:P_gg}
    P_{gg}(k,z) = \sum_i \sum_j b^L_i b^L_j P_{ij}(k,z) + P_\mr{SN},
\end{equation}

\begin{equation}
\label{eq:P_gm}
    P_{gm}(k,z) = \sum_i b^L_i P_{i\delta_m}(k,z),
\end{equation}

\noindent where $P_{ij}(k,z)$ comes from the contribution $\langle {\cal O}_i {\cal O}_j \rangle$ and $P_{i \delta_m}(k,z)$ from $ \langle {\cal O}_i \delta_m\rangle$, and the spectra depend only on the absolute value $k = |\bo k|$ as a result of isotropy. Here, we set $b_0 \equiv 1$ and define $P_{11}$ as the nonlinear matter power spectrum. The component $P_\mr{SN}$ arises from contributions to the galaxy field coming from processes below the typical host halo size, which we can assume to be uncorrelated with larger scales. We assume this ``stochastic contribution'' to be scale-independent in the range of scales relevant to this work. It is often identified with a Poisson shot-noise contribution. As described below, we keep it as a free parameter in the likelihood analyses.

We can apply Lagrangian perturbation theory \cite{Chen_2020} to calculate the displacement field $\bo \Psi$ and the power spectrum components $P_{ij}(k,z)$. In practice, we use the one-loop results provided by {\tt Velocileptors} \cite{Chen_2020, Chen_2021}. We neglect cubic Lagrangian bias parameters entering the one-loop $P(k)$ through the term $P_{13}(k)$, as they have been shown to be degenerate with lower-order parameters \cite{Abidi_2018, LSSTDarkEnergyScience:2023qfp}.

In the next subsection, we use $P_{gg}(k,z)$ and $P_{gm}(k,z)$ to define the data vectors and corresponding covariance matrices to appear in our analyses.

\subsection{Angular power spectra of galaxy clustering and galaxy-galaxy lensing}
\label{sec:models}

We are interested in the angular power spectra of galaxy clustering and galaxy-galaxy lensing. As conventional, we refer to lens galaxies as those responsible for the weak gravitational lensing of the so-called source galaxies. The relevant angular spectra can be obtained from $P_{gg}(k)$ and $P_{gm}(k)$, in the Limber approximation \cite{LoVerde_2008}, through:

\begin{equation}
\label{eq:limber_gg}
    C^{ij}_{gg}(\ell) = \int_0^\infty \mr d\chi \frac{\phi^i(\chi)\phi^j(\chi)}{\chi^2} P_{gg}\left( \frac{\ell}{\chi}, z(\chi) \right),
\end{equation}

\begin{equation}
\label{eq:limber_gk}
    C^{ij}_{g\kappa}(\ell) = \int_0^\infty \mr d\chi \frac{\phi^i(\chi)g^j(\chi)}{\chi^2} P_{gm}\left( \frac{\ell}{\chi}, z(\chi) \right),
\end{equation}

\noindent where $\phi^i(\chi)$ and $g^i(\chi)$, corresponding to redshift bin $i$, are the selection functions for clustering and lensing, respectively, and $\chi$ is the comoving radial distance to redshift $z$. Specifically, $\phi^i(\chi)$ is the radial density of galaxies in the $i$-th bin of the lens sample. The lens efficiency $g^i(\chi)$ in a flat cosmology is given by:

\begin{equation}
    g^i(\chi) = \frac{3 \Omega_m H_0^2}{2 c^2 a(\chi)} \int_0^\infty \mr dz \, n^i(z) \frac{(\chi'(z)-\chi)\chi}{\chi'(z)} \Theta(\chi'(z)-\chi),
\end{equation}

\noindent where $n^i(z)$ is the radial density of galaxies in the $i$-th bin of the source sample, $\Omega_m$ is the matter-density parameter, $c$ is the speed of light, $a(\chi)$ is the expansion scale factor as a function of $\chi$, $H_0$ is the present-day Hubble parameter, and $\Theta(\chi'(z) - \chi)$ is the Heaviside step function \cite{Krause_2017}.

While we assume the galaxy density field is non-Gaussian, we consider only the disconnected contribution to the covariance matrix. This is done to isolate the impact of including theoretical error in that matrix. The covariance between two angular power spectra is \cite{Hu_2004}:

\begin{equation}
\label{eq:cov_d}
\begin{aligned}
    \mathrm{Cov}\bigg[C^{ij}_{AB}(\ell_1), C^{kl}_{CD}(\ell_2)\bigg] &= \frac{4\pi\delta_{\ell_1\ell_2}}{\Omega_s(2\ell_1+1)\Delta\ell_1} \times \\
    &\bigg[(C^{ik}_{AC}(\ell_1)+\delta_{ik}\delta_{AC}N^i_A) (C^{jl}_{BD}(\ell_2)+\delta_{jl}\delta_{BD}N^j_B) +\\
    &(C^{il}_{AD}(\ell_1)+\delta_{il}\delta_{AD}N^i_A)(C^{jk}_{BC}(\ell_2)+\delta_{jk}\delta_{BC}N^j_B)\bigg].
\end{aligned}
\end{equation}    

\noindent Here, uppercase letters $A$ to $D$ represent the probes (either $g$, galaxy count, or $\kappa$, lensing) and indices $i$ to $k$ identify the redshift bin. The parameter $\Omega_s$ denotes the survey area. We choose a value consistent with the LSST: $\Omega_s/4\pi = 0.4$. The symbol $\Delta\ell$ denotes the length of $\ell$ binning. The expression above implies, for example, that there is nonzero covariance between spectra projected in different redshift bins, as well as between different spectra ($C_{gg}$ and $C_{g\kappa}$) defined in the same redshift bin. 

The parameters $N^i_A$ are probe-specific noise terms, among which we are concerned with shot noise and shape noise. When necessary, we assume the former to be given by $N^i_\mr{g} = 1/\Bar n_\mr{lens}^i$, where $\Bar n_\mr{lens}^i$ is the mean galaxy density in the $i$th lens bin \cite{Krause_2017}. Shape noise is $N^i_\kappa = \sigma^2_\epsilon / \Bar n^i_\mr{source}$, where $\Bar n^i_\mr{source}$ is the mean galaxy density in the $i$th source bin and $\sigma_\epsilon$ is the rms shear arising from the intrinsic ellipticity of galaxies and measurement noise \cite{Hu_2004}. We consider $\sigma_\epsilon = 0.26$ for simulated data vectors (subsection \ref{sec:lik_analysis_sdv}), and the DESC Bias Challenge adopts $\sigma_\epsilon = 0.28$ (subsection \ref{sec:lik_analysis_abac}) \cite{LSSTDarkEnergyScience:2023qfp}. In practice, we include these noise terms in the angular spectra $C^{ii}_{gg}(\ell)$ and $C^{ii}_{\kappa\kappa}(\ell)$ themselves and not explicitly in the covariance matrices. We specify the configurations of redshift binning and $\ell$-binning used in this work in section \ref{sec:lik_analysis}.

Having specified the Gaussian covariance matrix of our observables, we turn to the modeling of the theoretical error covariance in the next section.

%% file: TE_models.tex
\section{Models of theoretical-error covariance}
\label{sec:envelopes}

In this section, we present the models of theoretical-error covariance adopted in this work. We begin, as suggested in \cite{baldauf2016lss}, by writing the covariance matrix of theoretical error as:

\begin{equation}
\label{eq:cov_e}
    \mr{Cov}_e^{\ell\ell'} = E_{\ell} E_{\ell'} \rho_{\ell\ell'}.
\end{equation}

\noindent  The function $E_\ell$, named the \textit{envelope}, defines the diagonal of the covariance matrix, hence the variance of theoretical error. It can be viewed as the overall size of theoretical uncertainty on a given mode $\ell$. The coefficients $\rho_{\ell\ell'}$ specify the correlation between errors at different modes. We set $\rho_{\ell\ell} \equiv 1$ for all $\ell$. 

A 2$\times$2-point analysis depends on the angular spectra $C^{gg}_\ell$ and $C^{g\kappa}_\ell$, and we must calibrate the theoretical error associated with these two spectra. We have, therefore, three components of TE covariance with the form \eqref{eq:cov_e}. They are:

\begin{equation}
\begin{aligned}
\label{eq:cov_e_specific}
    \mr{Cov}_{gg,gg}^{\ell\ell'} &= E^{gg}_{\ell} E^{gg}_{\ell'} \rho_{\ell\ell'} , \\
    \mr{Cov}_{g\kappa,g\kappa}^{\ell\ell'} &= E^{g\kappa}_{\ell} E^{g\kappa}_{\ell'} \rho_{\ell\ell'} , \\
    \mr{Cov}_{gg,g\kappa}^{\ell\ell'} &= E^{gg}_{\ell} E^{g\kappa}_{\ell'} \rho_{\ell\ell'} ,
\end{aligned}
\end{equation}

\noindent where $E^{gg}_\ell$ and $E^{g\kappa}_\ell$ are the envelopes associated with $C^{gg}_\ell$ and $C^{g\kappa}_\ell$, respectively. Any symmetric matrix can be written as in eq. \eqref{eq:cov_e}, but eq. \eqref{eq:cov_e_specific} contains modeling choices. We assume a common function $\rho_{\ell\ell'}$ for all three matrices. We also suppose the diagonal of the $gg$-$g\kappa$ matrix can be written as $E^{gg}_{\ell} E^{g\kappa}_\ell$. This expresses the approximation: $\sigma^2_{XY} \simeq \sigma_X \sigma_Y$, where $X$ and $Y$ are random variables with covariance $\sigma^2_{XY}$ and standard deviations $\sigma_X$ and $\sigma_Y$. This has the advantage of permitting the entire TE covariance to be determined from the three functions $E^{gg}_\ell$, $E^{g\kappa}_\ell$, and $\rho_{\ell\ell'}$ alone. The same considerations apply to the TE covariance across redshift bins, as it can't be rigorously calculated from the standard deviations (envelopes) of the error in each bin. In light of the relatively large separation between our redshift bins (section \ref{sec:lik_analysis}), we set these covariances to zero, an approximation which \cite{Chudaykin_2021, Moreira_2021} found satisfactory. We test these assumptions by applying the resulting TE matrix to mock likelihood analyses, as described in sections \ref{sec:lik_analysis} and \ref{sec:results}.

In the following, we describe the modeling of the envelopes and correlation coefficients.

\subsection{Envelope}

The envelope is the standard deviation of theoretical error. It captures the uncertainty in the modeling of the galaxy angular power spectra due to contributions from neglected higher-order bias operators. Cosmological simulations are useful tools to characterize the statistical distribution of the coefficients associated with such operators. Since theoretical error arises from neglecting their contributions, the inferred distributions of bias parameters can be used to assess the theoretical error in analyses that use only linear bias.

We can derive the standard deviation of the error from a collection of angular spectra expressing a range of plausible bias conditions. This procedure was adopted by \cite{Moreira_2021} to calculate the uncertainty on baryonic-feedback models, in this case from a set of 13 hydrodynamical simulations assumed to be representative of the feedback process. Likewise, \cite{Chudaykin_2021} used LasDamas N-body simulations \cite{lasdamas} to quantify the uncertainty on the one-loop perturbative model of dark matter and galaxy density.

In this work we use the results of \cite{Barreira_2021}, based on fitting an EFT bias model to a set of IllustrisTNG simulations. This produces statistical relations between the higher-order parameters $b_2$ and $b_{s^2}$ and the linear parameter $b_1$. The best-fit values of $b_2$ and $b_{s^2}$ are:

\begin{equation}
\begin{aligned}
\label{eq:b2_bs}
    b_2 &= 0.24 \, b_1^3 + 1.12 \, b_1^2 - 0.44 \, b_1 - 0.34 , \\
    b_{s^2} &= -0.28 \, b_1 + 0.09 ,
\end{aligned}
\end{equation}

\noindent with standard deviations of

\begin{equation}
\label{eq:sigma_b2_bs}
    \sigma_2 = 0.44, \quad \sigma_{s^2} = 0.22 ,
\end{equation}

\noindent which are independent of $b_1$. Since the authors quote these relations in the Eulerian frame, we use the coevolution relations at second order \cite{Abidi_2018} to express them in the Lagrangian frame, which is adopted in this work. These are:

\begin{equation}
\label{eq:euler_lagrange_conv}
\begin{aligned}
    b_1^\mr{(E)} &= b_1^\mr{(L)} + 1 , \\
    b_2^\mr{(E)} &= \frac{4}{21} b_1^\mr{(L)} + \frac{1}{2} b_2^\mr{(L)} , \\
    b_{s^2}^\mr{(E)} &= -\frac{2}{7} b_1^\mr{(L)} ,
\end{aligned}
\end{equation}

\noindent which can be inverted to obtain $b_i^\mr{(L)}$ in terms of $b_j^\mr{(E)}$.

As to the relation between the best-fit $b_{\nabla^2\delta}$ and $b_1$, which is not provided by \cite{Barreira_2021}, we use a result of \cite{Zennaro_2022} based on BACCO N-body simulations \cite{Angulo_2021} and a subhalo-abundance matching technique for the creation of galaxy catalogs. The relation thus found is:

\begin{equation}
\label{eq:bnabla}
    b_{\nabla^2\delta} = 0.2298 \, b_1^3 - 2.096 \, b_1^2 + 0.7816 \, b_1 - 0.1545 ,
\end{equation}

\noindent already in the Lagrangian frame.

The authors of \cite{Zennaro_2022} report 100\% of their $b_{\nabla^2\delta}$ samples in the interval $[b_{\nabla^2\delta}^\mathrm{best-fit} - 5, \\b_{\nabla^2\delta}^\mathrm{best-fit} + 8]$. To interpret that range in terms of a Gaussian distribution, we assume it equal to two times the $3\sigma$ range, or the 99.78\% confidence interval, which implies:

\begin{equation}
\label{eq:sigma_bnabla}
    \sigma_{\nabla^2\delta} = 2.17 .
\end{equation}

We construct the envelope of theoretical error by sampling from the aforementioned distributions of bias parameters. For this purpose we generate a set of $10^4$ angular power spectra of galaxy clustering and galaxy-galaxy lensing. We compute the underlying 3D power spectra $P_{gg}(k,z)$ and $P_{gm}(k,z)$ in the real-space Lagrangian frame with {\tt Velocileptors} \cite{Chen_2020}, at one-loop order. These spectra contain the parameters $b_2$, $b_{s^2}$, and $b_{\nabla^2\delta}$ sampled from relations \eqref{eq:b2_bs} and \eqref{eq:bnabla} and Gaussian uncertainties \eqref{eq:sigma_b2_bs} and \eqref{eq:sigma_bnabla}. We assume these higher-order parameters to be uncorrelated. To convert 3D tracer spectra to the projected statistics of tomographic galaxy surveys we use the DESC Core Cosmology Library ({\tt CCL}) \cite{Chisari_2019}. 

Specifically, we generate a simulated data vector in the ensemble as follows:
\begin{enumerate}
    \item Choose a linear bias $b_1$. This must match, approximately, the $b_1$ value of the galaxy sample to which the method will be applied. In this case we estimate $b_1$ by applying a scale cut at $\kmax = 0.08 \hmpc$.
    \item Draw values for $b_2$, $b_{s^2}$, and  $b_{\nabla^2\delta}$ from \cref{eq:b2_bs,eq:sigma_b2_bs,eq:bnabla,eq:sigma_bnabla}.
    \item Construct $P_{gg}(k,z)$ and $P_{gm}(k,z)$ with {\tt Velocileptors}.
    \item Construct $C^{ij}_{gg}(\ell)$ and $C^{ij}_{g\kappa}(\ell)$ from eqs. \eqref{eq:limber_gg} and \eqref{eq:limber_gk}.
\end{enumerate}

From each sampled $C^{ij}_{gg}(\ell)$ and $C^{ij}_{g\kappa}(\ell)$ we define a corresponding theoretical error as its residual with respect to the prediction of a linear-bias model. Here, we do not add shot noise either to the simulated data vector or to the linear-bias prediction, supposing all the theoretical error comes from the bias parameters.

We generate the envelope from this ensemble of theoretical errors. Specifically, for each simulated data vector we determine the best-fit value $b_1^\mr{bf}$ in a model containing only linear bias. We give more details on this fit below. The error vector is then 

\begin{equation}
\label{eq:te_residual}
    e_\ell = d_\ell (b_1,b_2, b_{s^2},  b_{\nabla^2\delta} ) - t_\ell(b_1^\mr{bf}) ,
\end{equation}

\noindent where $d_\ell = \{C^{ij}_{gg}(\ell),C^{ij}_{g\kappa}(\ell)\}$ is the simulated data vector and $t_\ell = \{C^{ij (\mr{th})}_{gg}(\ell),C^{ij(\mr{th})}_{g\kappa}(\ell)\}$ is the theoretical vector computed with linear bias only.

We determine the envelope from the ensemble's statistics. Two possibilities are adopted: the standard deviation of theoretical errors (named the ``1 sigma'' envelope) and the 95th percentile of absolute theoretical errors (named ``95\%''). The latter choice, which is approximately two standard deviations, represents a conservative assumption on the uncertainty. We find likelihood analyses with this envelope to be more robust under variations of the data vector, as will be seen in subsection \ref{sec:results_sdv}. From the same ensemble we determine the mean-error vector entering the modified likelihood \eqref{eq:mod_likelihood}.

\begin{figure}[tbp]
\centering % \begin{center}/\end{center} takes some additional vertical space
\includegraphics[width=0.45\textwidth,clip]{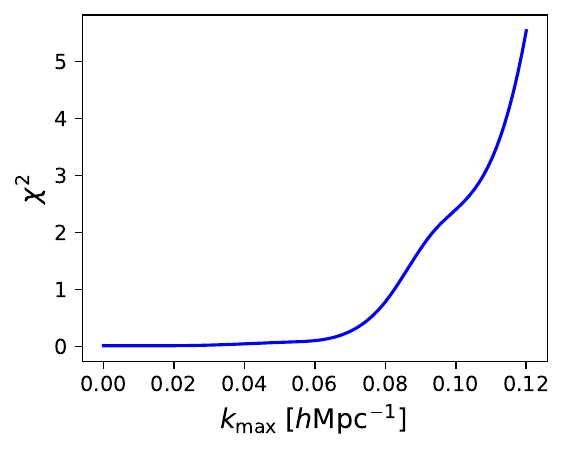}
\hfill
\caption{\label{fig:chi_square} $\chi^2$ deviation as a function of $\kmax$ in an estimation with the linear-bias model. Here we use the noiseless data vectors studied in the DESC Bias Challenge \cite{LSSTDarkEnergyScience:2023qfp}, to be described in subsection \ref{sec:lik_analysis_abac}.}
\end{figure}

The fitting step \eqref{eq:te_residual} requires more explanation. Since we must calculate the best-fit spectrum $t_\ell$ on a range of $\ell$ where the linear-bias model is reliable, this entails a choice of $\ell_\mr{max}$. We could determine it from a series of scale cuts (at various $\ell_\mr{max}$ values) applied, for example, to an estimation of cosmological parameters $\Omega_c$ and $\sigma_8$. As this would be computationally expensive, we use instead the $\chi^2$-based criterion adopted in \cite{Chudaykin_2021}:

\begin{enumerate}
    \item Given a typical simulated data vector generated with the procedure above (i.e., containing nonlinear bias), we fit a linear-bias model to it by varying $b_1$ and keeping the cosmology fixed. Since only a bias parameter is varied, this step is computationally inexpensive. We repeat it for various values of $\kmax$, in each case determining $\ell_\mr{max}$ from the relation $\ell_\mr{max} = \kmax \chi(z_p)$, where $\chi(z_p)$ is the radial distance to the maximum of the redshift bin. For each such $\kmax$, the $\chi^2$ of the fit is registered. Hence this procedure yields a function $\chi^2(\kmax)$.
    \item This function is typically increasing, as exemplified in figure \ref{fig:chi_square}. At scales where linear bias is no longer a sufficiently accurate model, we expect it to begin growing steeply. Any value of $\kmax$ below that point is suitable for an estimation with linear bias, and we checked that the envelope doesn't depend strongly on this choice. It depends, however, on the data vector used in the fit. We choose $\kmax = 0.06 \hmpc$ for the case of simulated data vectors (to be described in subsection \ref{sec:lik_analysis_sdv}) and $\kmax = 0.07 \hmpc$ for the case of data vectors studied in the DESC Bias Challenge (described in subsection \ref{sec:lik_analysis_abac}). The fit is performed with $\ell_\mr{min} = 20$ to agree with LSST's prescriptions \cite{LSSTDarkEnergyScience:2023qfp}.
\end{enumerate}

\begin{figure}
    \centering
    \begin{subfigure}[t]{\textwidth}
        \centering
        \includegraphics[width=\linewidth]{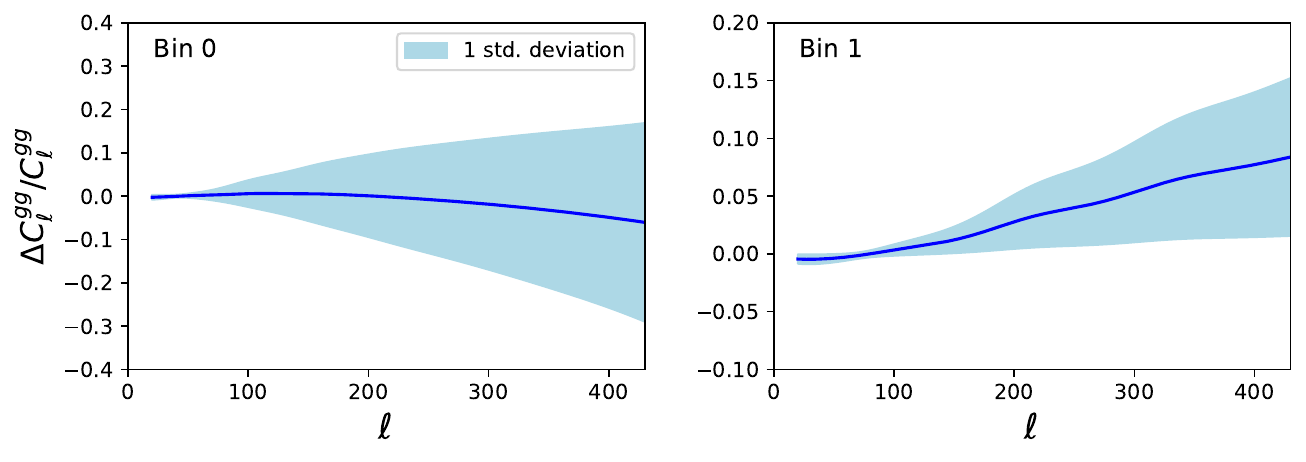} 
    \end{subfigure}
    \vspace{1cm}
    \begin{subfigure}[t]{\textwidth}
        \centering
        \includegraphics[width=\linewidth]{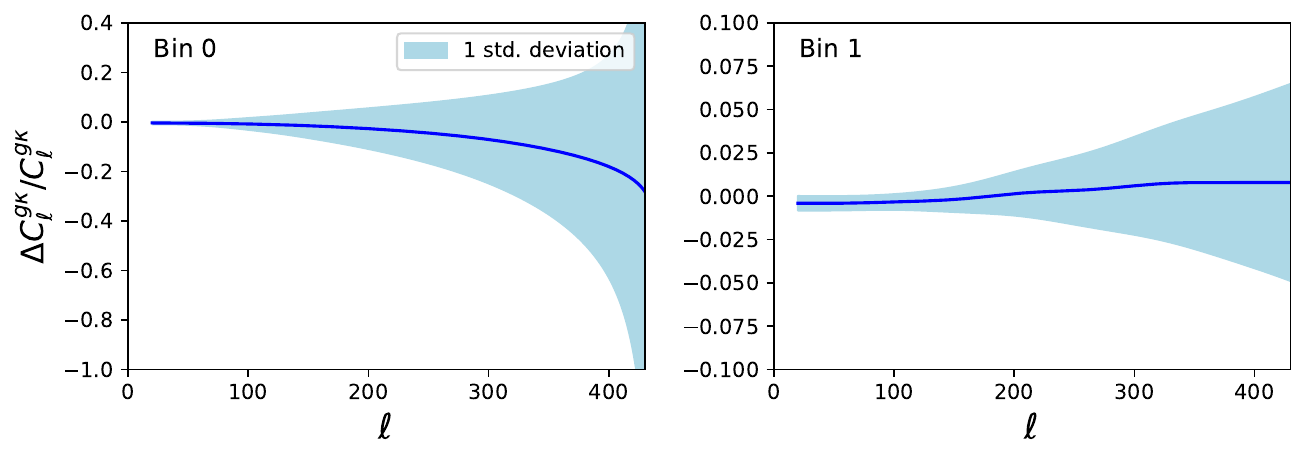} 
    \end{subfigure}
    \caption{\label{fig:rel_errors} Relative errors on $C^{gg}_\ell$ and $C^{g\kappa}_\ell$ in an analysis with the simulated data vectors described in subsection \ref{sec:lik_analysis_sdv}. The lens redshift bins have maxima at $z = 0.36$ (bin 0) and $z = 0.84$ (bin 1). The source bins have maxima at $z = 0.34$ (bin 0) and $z = 0.86$ (bin 1).}
\end{figure}

The range of relative errors on $C^{ij}_{gg}(\ell)$ and $C^{ij}_{g\kappa}(\ell)$ obtained with this prescription is shown in figure \ref{fig:rel_errors} for two redshift bins. As expected, the uncertainty increases as the spatial scale decreases. We remark also that the mean error deviates from zero, which justifies its inclusion in the modified likelihood \eqref{eq:mod_likelihood}. 

\subsection{Correlation coefficients}

We now discuss the modeling of the correlation coefficients $\rho_{\ell\ell'}$ in the TE covariance matrix \eqref{eq:cov_e}. These determine the correlation of theoretical errors at different $\ell$-modes. We adopt for the form of $\rho_{\ell\ell'}$ a well-behaved function which is symmetric under $\ell \leftrightarrow \ell'$:

\begin{equation}
\label{eq:rho}
    \rho_{\ell\ell'}= \mathrm{exp}\left[- \frac{(\ell-\ell')^2}{2\Delta \ell^2}\right] .
\end{equation}

\noindent This function has been applied with success to analyses of baryonic feedback \cite{Moreira_2021} and higher-loop contributions to the matter power spectrum \cite{baldauf2016lss}. In the above, $\Delta \ell$ is the typical coherence length of theoretical error. The condition $\Delta \ell \neq 0$ expresses the idea that errors at neighboring $\ell$ values must be similar, i.e., correlated. 

The coherence length can't be determined from first principles by any available method. If we had detailed information on the probability distribution of bias parameters (on their correlations, for example), we could attempt to determine the entire TE covariance matrix, including the effective correlation length, from the ensemble of theoretical errors. This means extending to the whole TE matrix the procedure described above for the computation of its diagonal. In this case the ansatz \eqref{eq:rho} would be unnecessary. It is unlikely, however, that such detailed information will be available in all cases where the TE method is to be applied. 

For the value of $\Delta \ell$, therefore, we test a few possibilities. Our first choice is $\Delta \ell = 0$, which produces a diagonal matrix and is appealing for its simplicity. This has the disadvantage, however, of making the TE covariance dependent on $\ell$-binning. Moreover, the assumption that theoretical errors on neighbouring $\ell$ values are uncorrelated -- which implies the corresponding data points are independent -- can introduce spurious precision in the analysis. This expectation will be borne out in the results of section \ref{sec:results}.

A second possibility, adopted by \cite{baldauf2016lss} and \cite{Chudaykin_2021}, is to assume that the coherence length is large enough to ensure correlation at the typical BAO scale, $\Delta k_\mathrm{BAO} = 0.05 \hmpc$. For instance, the value $\Delta k = 0.1 \hmpc$ is chosen in \cite{Chudaykin_2021}. As the BAOs are not as distinct a feature of angular power spectra as they are of three-dimensional spectra, however, it is not clear whether $\Delta k_{\rm BAO}$ is a relevant length. Nevertheless, we include it in our tests to establish a baseline with respect to previous literature.

\begin{figure}[tbp]
\centering % \begin{center}/\end{center} takes some additional vertical space
\includegraphics[width=0.5\textwidth,clip]{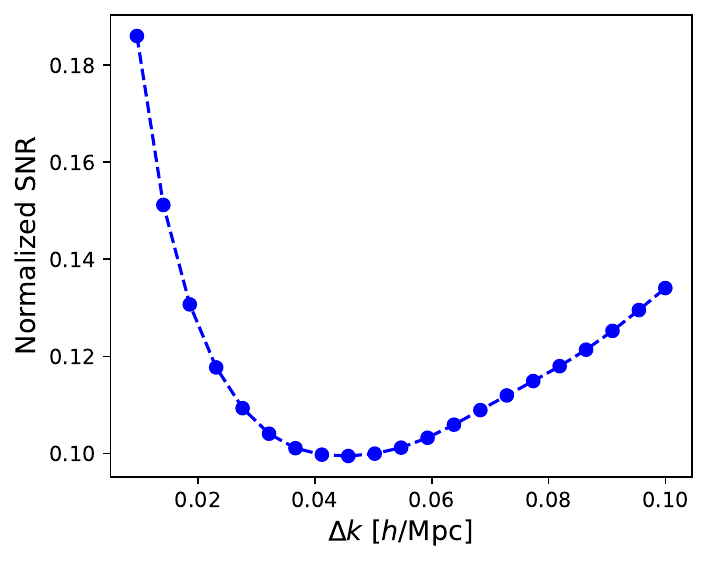}
\hfill
\caption{\label{fig:snr_vs_deltak}
 $\mr{SNR}$ as a function of $\Delta k$, computed with eq. \eqref{eq:snr} and normalized by its value at $\Delta k = 0$. Here we use data vectors studied in the DESC Bias Challenge \cite{LSSTDarkEnergyScience:2023qfp} and described in subsection \ref{sec:lik_analysis_abac}. We select for test that value of $\Delta k$ which minimizes this function, approximately $\Delta k = 0.05 \hmpc$.}
\end{figure}

Before introducing a third test value for $\Delta \ell$, we remark that choosing too \emph{large} a coherence length can also produce spurious precision. This is because theoretical vectors with relatively sharp variations across $\Delta \ell$ will be penalized by the likelihood function, which implies a reduction of the region in parameter space where the likelihood is dense, i.e., an increase in precision. We illustrate this situation in appendix \ref{sec:appendix} by means of a toy model and with simulation-based data vectors in figure \ref{fig:snr_vs_deltak}.

In this scenario, one conservative option is to minimize precision as a function of the coherence length. This is to safely avoid the possibility of any spurious precision being connected with $\Delta \ell$, and it can be justified in that no rigorous method is available to determine that length from first principles. We thus introduce the signal-to-noise ratio (SNR):

\begin{equation}
\label{eq:snr}
    \mr{SNR}(\Delta k) = \bo t \, \mr{Cov^{-1}}_T(\Delta k) \, \bo t^\mr{T} ,
\end{equation}

\noindent where $\Delta k = \Delta\ell(z_p)/\chi(z_p)$, $\bo t = \{ C_{gg}(\ell), C_{g\kappa}(\ell) \}$ is a theoretical vector, and $\mr{Cov}_T$ is the augmented covariance matrix, which depends on the coherence length $\Delta k$. To evaluate the SNR as a function of $\Delta k$, we first compute the theoretical vector in the linear-bias model, choosing the $b_1$ value that best fits the full-bias data vector. The calculation of $\mr{SNR}(\Delta k)$ is computationally inexpensive, as the theoretical vector is fixed and only the exponent in eq. \eqref{eq:rho} is varied.

We show its typical form in figure \ref{fig:snr_vs_deltak}. As anticipated above, the function $\mr{SNR}(\Delta k)$ increases for large $\Delta k$, which expresses the gain in precision caused by the exclusion of theoretical vectors with sharp variations across $\Delta \ell = \chi(z_p) \Delta k$. We thus also adopt a value of $\Delta k$ which minimizes this function, approximately $\Delta k = 0.05 \hmpc$. We have checked that this minimum is nearly the same for typical data vectors issued from the two data sets described in section \ref{sec:lik_analysis}.\footnote{An approximate value of the minimum is sufficient for our purposes, i.e., to avoid the spurious gain in precision associated with very large or very small $\Delta k$.} The three $\Delta k$ values chosen for the mock likelihood analyses are, therefore, 0, 0.05, and $0.10 \hmpc$.

We turn to a description of the data vectors and other aspects of the likelihood analyses in the next section.

%% file: lik_analysis.tex
\section{Likelihood analysis}
\label{sec:lik_analysis}
We now discuss the methodology used to test the theoretical-error technique as applied to the uncertainty on galaxy bias. In particular, we define criteria for the comparison of scale cuts and TE based on accuracy, precision, and (when possible) robustness under variations of the data vector. We also test whether TE gives results independent of $\kmax$.

In order to focus on the effects on the nonlinear bias, we perform a simplified likelihood analysis with only two redshift bins. A more realistic study using 5 or 10 redshift bins, as prescribed by the LSST DESC \cite{LSSTDarkEnergyScience:2023qfp}, and including systematic effects such as magnification bias, intrinsic alignments, and photo-z uncertainties, can be performed in subsequent work if the simple treatment proves promising.

We adopt, as mock data vectors, angular power spectra of galaxy clustering and galaxy--galaxy lensing. Since higher-order bias parameters grow with $b_1$ (as shown in eqs. \eqref{eq:b2_bs} and \eqref{eq:bnabla}), and the latter grows with redshift, we expect the error due to nonlinear bias to similarly vary with redshift. We thus work with redshift bins well separated in the interval $\sim [0, 1]$, not to bias the analysis with either too weak or too strong a higher-order effect. 

We divide the mock data vectors in two sets:

\begin{itemize}
    \item Simulated data vectors: Spectra whose nonlinear bias parameters $\{b_2, b_{s^2}, b_{\nabla^2}\}$ are drawn from the distributions \eqref{eq:b2_bs}, \eqref{eq:bnabla}, \eqref{eq:sigma_b2_bs}, and \eqref{eq:sigma_bnabla} -- that is, from the same distributions that originate the envelopes and mean-error vectors --, corresponding to certain values of $b_1$. We consider three subsets of data vectors (``typical'', ``high'', and ``low'' bias) expressing different conditions of bias. 
    \item Mock LSST-like spectra used in the LSST Bias Challenge~\cite{LSSTDarkEnergyScience:2023qfp}. These are generated from galaxy--halo connections applied to the {\tt AbacusSummit} $N$-body simulation suite \cite{Garrison_2018}. We choose spectra corresponding to the ``HSC'' galaxy sample.
\end{itemize}

There are trade-offs with respect to these two sets of data vectors. The former, being sampled from a known probability distribution, may be generated with an arbitrary number of spectra and permits assessing the robustness of TE under variations of the data vector. As they are calculated from the perturbative bias expansion itself, however, they can't be used realistically to study performance on scales beyond the valid regime $k \lesssim 0.3 \hmpc$ \cite{Desjacques:2016bnm, LSSTDarkEnergyScience:2023qfp}. The second set, despite consisting of a single data vector, is valid towards very small scales, $k \sim 1 \hmpc$, and allows us to reliably assess the $\kmax$-dependence of TE.

We give more details on the analyses with these two sets below.

\subsection{Simulated data vectors}
\label{sec:lik_analysis_sdv}

We use three simulated data vectors differing in conditions of nonlinear galaxy bias. This is to avoid that we misrepresent the effectiveness of one method because of an exceptionally favorable (or unfavorable) data vector -- e.g., a scale cut is expected to perform better in cases of weak nonlinear bias. That this care must be taken is exemplified by the results of \cite{Moreira_2021}, where the inclusion of theoretical error led to significant gains in situations of moderate-to-strong baryonic feedback, but was ineffectual (although harmless) when applied to models where these effects were particularly weak -- e.g., the MassiveBlack-II model. 

The three data vectors are defined as:

\begin{enumerate}
    \item Typical bias: The parameter $b_1$ has values 0.7 and 1.0 in bins 0 and 1, respectively. These values (rather, their counterparts in the Eulerian frame) are used in the DES Y3 analysis to stress-test scale cuts \cite{krause2021dark}. The parameter $b_2$ used in DES, and also in this work, is interpolated from the $b_2-b_1$ relation measured from catalogs of \texttt{redMagic}-like galaxies \cite{Rozo_2016}. The coefficient $b_{s^2}$ is computed from $b_1$ through the coevolution relation (in the Eulerian frame) $b_{s^2} = -4(b_1-1)/7$. The parameter $b_{\nabla^2}$, which is absent in the DES analysis, is computed from $b_1$ from the best-fit value \eqref{eq:bnabla}.
    \item High bias: All nonlinear-bias parameters (i.e., $b_2$, $b_{s^2}$, and $b_{\nabla^2}$) are 0.7-$\sigma$ deviated upwards, according to the standard deviations \eqref{eq:sigma_b2_bs} and \eqref{eq:sigma_bnabla}. The corresponding spectra $C^{gg}_\ell$ and $C^{g\kappa}_\ell$ are nearly one standard deviation larger than the central spectra -- that is, those computed with the central parameters \eqref{eq:b2_bs} and \eqref{eq:bnabla}.
    \item Low bias: All nonlinear-bias parameters are 0.7-$\sigma$ deviated downwards. The spectra $C^{gg}_\ell$ and $C^{g\kappa}_\ell$ are nearly one standard deviation lower than the central spectra.
\end{enumerate}

\begin{figure}
    \centering
    \begin{subfigure}[t]{\textwidth}
        \centering
        \includegraphics[width=\linewidth]{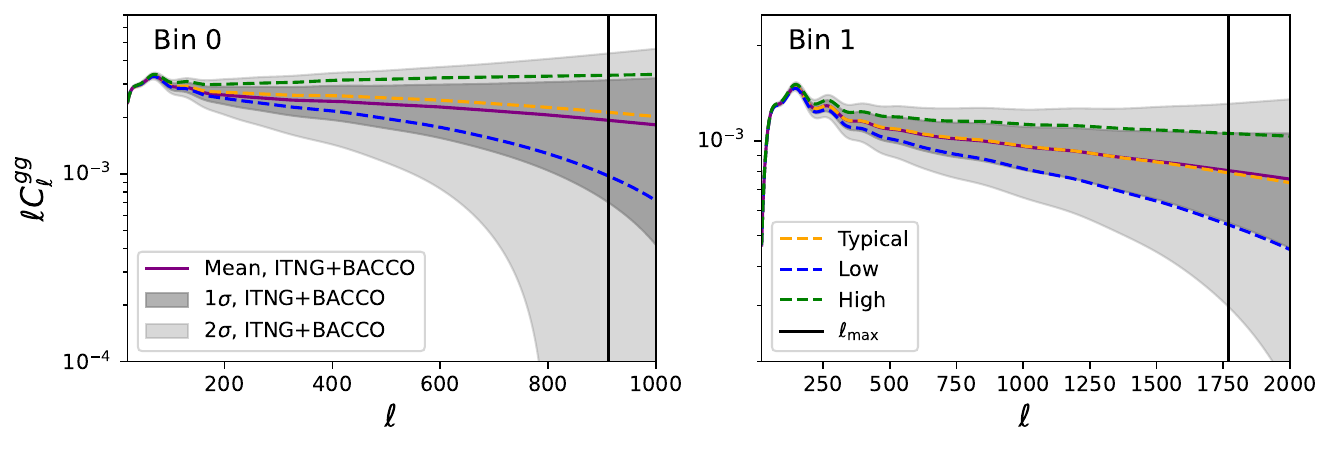} 
    \end{subfigure}
    \vspace{1cm}
    \begin{subfigure}[t]{\textwidth}
        \centering
        \includegraphics[width=\linewidth]{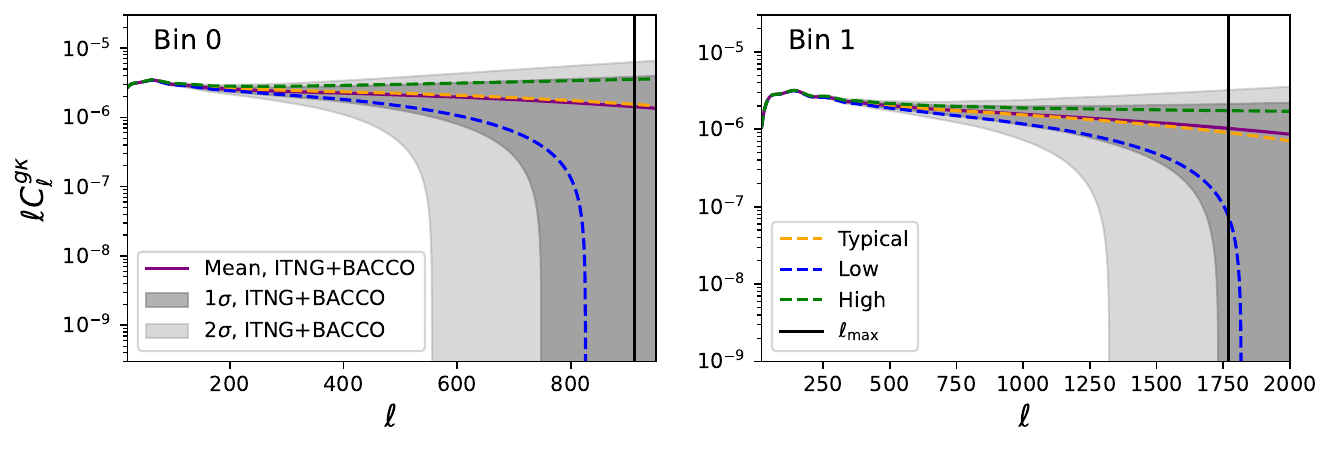} 
    \end{subfigure}
    \caption{\label{fig:sim_data} Data vectors of galaxy clustering and galaxy-galaxy lensing differing in the strength of nonlinear galaxy bias, for lens redshift bins with peaks at $z = 0.36$ (bin 0) and $z = 0.84$ (bin 1). The shaded areas represent the 1-sigma and 2-sigma ranges corresponding to the simulation-based relations \eqref{eq:b2_bs}, \eqref{eq:sigma_b2_bs}, \eqref{eq:bnabla}, and \eqref{eq:sigma_bnabla} (``ITNG+BACCO''). The negative values attained by $C^{g\kappa}_\ell$ at large $\ell$ modes, seen above for redshift bin 0, are a result of applying the perturbative model beyond $\kmax \approx 0.3 \hmpc$. For reliable predictions at high $\ell$ modes, we use data vectors based on the {\tt AbacusSummit} N-body simulation suite (subsection \ref{sec:lik_analysis_abac}).}
\end{figure}

The three sets of data vectors described above are shown in figure \ref{fig:sim_data}, where we also present the mean spectra and the standard deviations corresponding to the simulation-based distributions described in section \ref{sec:envelopes} (``ITNG+BACCO''). The negative excursion of $C^{g\kappa}_\ell$ at large $\ell$ values, seen in the figure for the ``low bias'' data vector at bin 0, is a symptom of extending the perturbative model beyond its natural limit $\kmax \approx 0.3 \hmpc$. Here we use $\kmax = 1 \hmpc$. Specifically, the $\nabla^2 \delta$ term has a negative coefficient and dominates the signal in the ``low bias'' case. This can be taken as a feature of stress-testing, as it potentially worsens the results with TE but doesn't affect the scale cut, which is applied at much smaller $\kmax$. We turn to data vectors based on the {\tt AbacusSummit} N-body simulation suite for reliable predictions at high $k$ modes (subsection \ref{sec:lik_analysis_abac}). 

The bias parameters for the three sets of data vectors are specified in table \ref{tab:fid_bias}. We note that the ``high bias'' and ``low bias'' terminology refers to the signed value of the coefficients, not to their (absolute) magnitudes. For example, for bin 1, the modulus $|b_{s^2}|$ is larger in the ``low bias'' than in the ``high bias'' case (table \ref{tab:fid_bias}). In this sense, both the ``low bias'' and ``high bias'' vectors can be said to represent cases of nonlinear bias stronger than typical.

\begin{figure}[tbp]
\centering % \begin{center}/\end{center} takes some additional vertical space
\includegraphics[width=.75\textwidth,clip]{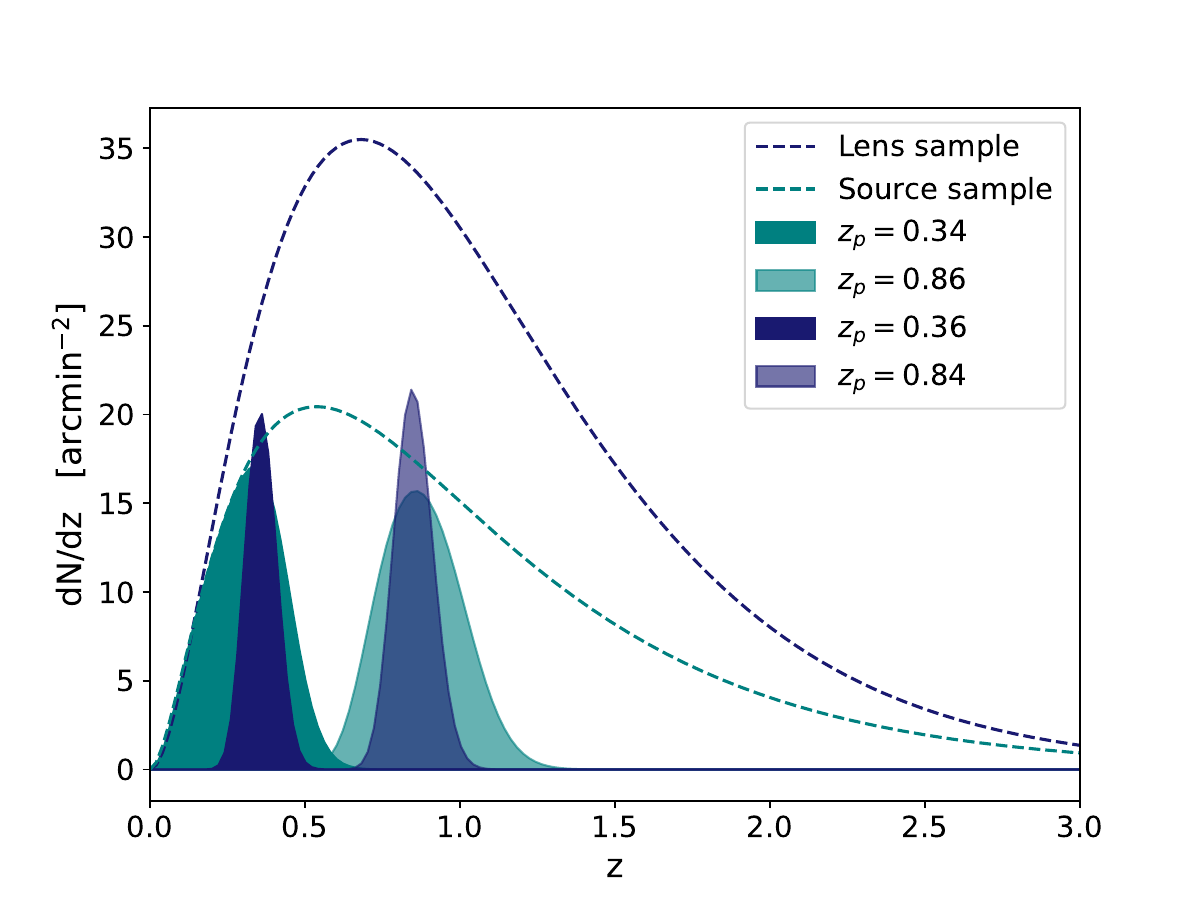}
\hfill
\caption{\label{fig:all_bins} Redshift bins of galaxy clustering and galaxy lensing selected for analysis in this work. The lens sample (red dashed curve) satisfies the distribution $\mr dn/\mr dz \propto z^2 \mathrm{exp}[-(z/z_0)^\alpha]$, with $z_0 = 0.28$ and $\alpha = 0.90$ and an overall normalization of 48 $\mathrm{arcmin}^{-2}$. The source sample (green dashed curve) satisfies $\mr dn/ \mr dz \propto z^2 \mathrm{exp}[-(z/z_0)^\alpha]$ with $z_0 = 0.11$ and $\alpha = 0.68$ and an overall normalization of 27 $\mathrm{arcmin}^{-2}$.}
\end{figure}

As to $\ell$-binning and redshift binning, we adopt the year-10 conventions of LSST-DESC \cite{thelsstdarkenergysciencecollaboration2021lsst}. These prescribe 20 $\ell$-bins logarithmically spaced between $20$ and $15000$. The tomographic analysis of galaxy clustering is segmented in 10 redshift bins on the interval $0.2 \le z \le 1.2$, spaced by 0.1 in photo-z. Of these, we select two for use in the estimations, with peaks at $z = 0.36$ and $z = 0.84$ (figure \ref{fig:all_bins}). The analysis of weak lensing is divided in 5 bins in that same interval, the photo-z limits being chosen so that each bin contains the same galaxy density. We choose source samples with maxima at $z = 0.34$ and $z = 0.86$ (figure \ref{fig:all_bins}). We select four
spectra as data vectors: $C^{00}_{gg}$, $C^{00}_{g\kappa}$, $C^{11}_{gg}$, and $C^{11}_{g\kappa}$. As will be seen in section \ref{sec:results}, this set permits clearly distinguishing TE and scale cut in terms of accuracy, precision, and robustness.\footnote{The spectrum $C^{01}_{g\kappa}$ has a higher intensity than $C^{00}_{g\kappa}$ and could have been used in its stead. However, we have found that the analyses with either $C^{01}_{g\kappa}$ or $C^{00}_{g\kappa}$ are equally informative, as measured by the total signal-to-noise ratios (SNRs). This is because $C^{01}_{g\kappa}$ has a stronger correlation with $C^{11}_{gg}$ and $C^{11}_{g\kappa}$, which lowers its contribution to the SNR.}

The lens sample satisfies the distribution $\mr dn/\mr dz \propto z^2 \mathrm{exp}[-(z/z_0)^\alpha]$, with $z_0 = 0.28$ and $\alpha = 0.90$ and an overall normalization of 48 $\mathrm{arcmin}^{-2}$. The photo-z scatter is $\sigma_z = 0.03(1+z)$. The source sample satisfies the same parametric distribution, now with $z_0 = 0.11$ and $\alpha = 0.68$, and an overall normalization of 27 $\mathrm{arcmin}^{-2}$. The photo-z scatter is $\sigma_z = 0.05(1+z)$ in this case.

The fiducial cosmology adopted is flat $\Lambda$CDM with $\Omega_c = 0.266$, $\Omega_b = 0.05$, $\sigma_8 = 0.831$, $h = 0.67$, and $n_s = 0.9645$. We take $\Omega_c$ and $\sigma_8$ as the only free cosmological parameters, whereas linear bias $b_1$ and shot noise at the two redshift bins are nuisance parameters. We assume wide flat priors on both cosmological and nuisance parameters, except for shot noise, which is sampled from a flat distribution in the range of $\pm 30\%$ of its Poissonian value~\cite{Kokron_2022_priors}. These priors are specified in table \ref{tab:priors_sdv}. We neglect the evolution of $b_1(z)$ within a redshift bin for these test vectors, an approximation that \cite{krause2021dark} has found adequate.

\input{priors_sdv}

Implicit in the considerations below is that theoretical errors on galaxy bias are fairly independent of the fiducial cosmology: the envelope is adjusted with a mock data set at the chosen cosmology to be later applied to data describing the actual universe. One commonly assumes the same independence when computing the $k_\mathrm{max}$ of a scale cut \cite{Tegmark_1997}.

We calculate the envelope and mean-error vector at the same cosmology used to generate the mock data vectors. In order to assess robustness of the envelope under variations of the fiducial cosmology, we also test this envelope and mean error against data vectors generated at two additional sets of cosmological parameters. Specifically, we vary $\Omega_c$ and $\sigma_8$ approximately within the 95\% confidence level of recent observational bounds \cite{Planck2018}. This is similar to the strategy adopted in \cite{Chudaykin_2021}.

% \usepackage{multirow}
\begin{table}[]
\centering
\begin{tabular}{|l|cccc|cccc|}
\hline
\multirow{2}{*}{Data vector} & \multicolumn{4}{c|}{$z_p = 0.34$ (lens)} & \multicolumn{4}{c|}{$z_p = 0.86$ (lens)} \\ \cline{2-9} 
                             & $b_1$  & $b_2$   & $b_{s^2}$  & $b_{\nabla^2}$  & $b_1$   & $b_2$  & $b_{s^2}$  & $b_{\nabla^2}$  \\ \hline
Typical bias                 & 0.700   & 0.190   & -0.200     & -0.556          & 1.000    & 0.620  & -0.280     & -1.239          \\ \hline
Low bias                    & 0.700   & -0.118  & -0.354     & -2.075          & 1.000    & 0.312  & -0.434     & -2.758          \\ \hline
High bias                  & 0.700   & 0.498   & -0.046     & 0.963           & 1.000    & 0.928  & -0.126     & 0.280          \\ \hline
\end{tabular}
\caption{Bias parameters defining the three pairs of simulated data vectors used in our analyses. ``Typical bias'' has the same set of $b_1$, $b_2$, and $b_{s^2}$ values as the vector used for validation in DES Year 3 \cite{krause2021dark}, when translated to the Lagrangian frame. We compute $b_{\nabla^2}$ through eq. \eqref{eq:bnabla}, given $b_1$. We obtain the parameters entering ``low bias'' and ``high bias'' from the first set by applying deviations of $-0.7\sigma_i$ and $+0.7 \sigma_i$, respectively, with $\sigma_i$ given in eqs. \eqref{eq:sigma_b2_bs} and \eqref{eq:sigma_bnabla}.}
\label{tab:fid_bias}
\end{table}

One of the aims of this study is to determine whether a calibration of $k_\mr{max}$ can be avoided in TE for the case of 2$\times$2-point tomographic analyses.\footnote{A choice of $\kmax$ is necessary for the construction of the envelope. As noted in section \ref{sec:envelopes}, however, this can be determined from a study of the function $\chi^2$ vs. $\kmax$ alone, where only one bias parameter is varied and the cosmology is kept fixed. Hence the procedure is less computationally expensive than the usual determination of $\kmax$ for scale cuts \cite{krause2021dark}.} If the envelope grows sufficiently fast with $\ell$, modes on small scales are suppressed automatically, which makes the specification of a $k_\mr{max}$ unnecessary. We deem an analysis with TE independent of $k_\mr{max}$ if:

\begin{itemize}
    \item For very-large $k_\mr{max}$, it recovers the fiducial cosmology with tension below or equal to 0.3$\sigma$. We take $k_\mr{max} = 1h \, \mr{Mpc}^{-1}$ for this test. This implies the method performs well under virtually any choice of $\kmax$ used in practice, likely much smaller than $1 \hmpc$. 
    \item The analysis is robust under changes of the galaxy sample. That is, when the data vector is varied from ``typical bias'' to ``low bias'' or ``high bias'', the best-fit parameters still deviate from the fiducial values with tension smaller than $\simeq 1\sigma$, also for $\kmax = 1\hmpc$. 
\end{itemize}

We use the following procedure to compare the results of the two methods. We adjust the $\kmax$ of the scale cut to produce the same tension with fiducial parameters observed with TE. This is done, in practice, by first running estimations with the scale cut at various values of $\kmax$. For comparison with a given TE configuration (e.g., a certain choice of envelope and coherence length), we select that $\kmax$ of scale cut which has produced similar tension with the fiducial cosmology.\footnote{An adjustment of scale cut to TE is more convenient than one of TE to the scale cut, since TE doesn't contain a continuous and adjustable parameter. As proposed here, the only free choices in this scheme are the envelope and the coherence length $\Delta k$, which are taken from a discrete and small set of options. Its $\kmax$ is kept fixed at $1 \hmpc$ to test $\kmax$-independence, whereas the $\kmax$ of the scale cut is freely adjustable.} We thus consider more effective whatever method yields higher precision, as assessed by the figure of merit:

\begin{equation}
    \mathrm{FoM} \equiv \frac{1}{\sqrt{\mathrm{det} [S_{\alpha\beta} / \theta_{\mathrm{fid},\alpha} \theta_{\mathrm{fid,\beta}}}]} ,
\end{equation}

\noindent where $S_{\alpha\beta}$ is the covariance matrix of the posterior, marginalized over the nuisance parameters $b_1^{(i)}$ and shot noise, and $\theta_{\mathrm{fid},\alpha}$ is the fiducial value of parameter $\theta_\alpha$, either $\Omega_c$ or $\sigma_8$. 

We evaluate the tension with the fiducial cosmology by measuring the marginalized 2D posterior at two points: its maximum and the point corresponding to fiducial parameters. We then translate the corresponding ratio as a sigma tension. We consider the analysis successful if it recovers the fiducial cosmology with a tension below or equal to 0.3$\sigma$. It will be convenient to define, as another measure of estimation error, the ``figure of bias'':

\begin{equation}
    \mathrm{FoB} \equiv \left[  \sum_{\alpha,\beta} (\bar{\theta}_\alpha - \theta_{\mathrm{fid},\alpha}) S^{-1}_{\alpha\beta} (\bar{\theta}_\beta - \theta_{\mathrm{fid},\beta})\right]^{1/2} ,
\end{equation}

\noindent where $\bar{\theta}_\alpha$ is the mean value of $\theta_\alpha$. This gives a measure of estimation error with respect to the overall uncertainty.

To check for projection effects, we run a separate analysis with a simulated data vector without nonlinear bias and a scale cut at $\kmax = 0.08 \hmpc$. As expected, the linear bias model is able to recover the fiducial cosmology with very low tension in the marginalized 2D posterior of $\sigma_8$ and $\Omega_c$. This indicates that projection effects are negligible for our parameter space. We give details in appendix \ref{sec:app_proj}.

\subsection{Data vectors used in the DESC Bias Challenge}
\label{sec:lik_analysis_abac}

For the second set of data vectors, we use galaxy samples based on the {\tt AbacusSummit} suite of N-body simulations \cite{Garrison_2018} and treated to angular power spectra in the DESC Bias Challenge \cite{LSSTDarkEnergyScience:2023qfp}. These are accurate, in principle, through $k \sim 1h \, \mr{Mpc}^{-1}$, hence convenient for tests of $k_\mr{max}$-independence by the criterion described above. Specifically, we use the ``maglim'' lens samples of the DESC Bias Challenge, whose parameters in the halo occupation distribution (HOD) are derived from Hyper Suprime-Cam (HSC) Y1 results presented in \cite{Nicola_2020}. The angular power spectra are calculated from the three-dimensional power spectra under the Limber approximation. Intrinsic alignments are not considered \cite{LSSTDarkEnergyScience:2023qfp}.

For the analysis of Abacus samples, we find models with constant $b_1(z)$ across a redshift bin to be inadequate, i.e., to introduce too large tensions with fiducial parameters even under aggressive scale cuts. We thus implement an evolution of the type $b_1(z) \sim 1/D(z)$ within the bin, which has the advantage of not requiring an extra parameter. This approximation has been found satisfactory for HSC galaxy samples \cite{Nicola_2020}, which form the basis for the maglim samples used in the DESC Bias Challenge \cite{LSSTDarkEnergyScience:2023qfp}.

\begin{figure}[tbp]
\centering % \begin{center}/\end{center} takes some additional vertical space
\includegraphics[width=.6\textwidth,clip]{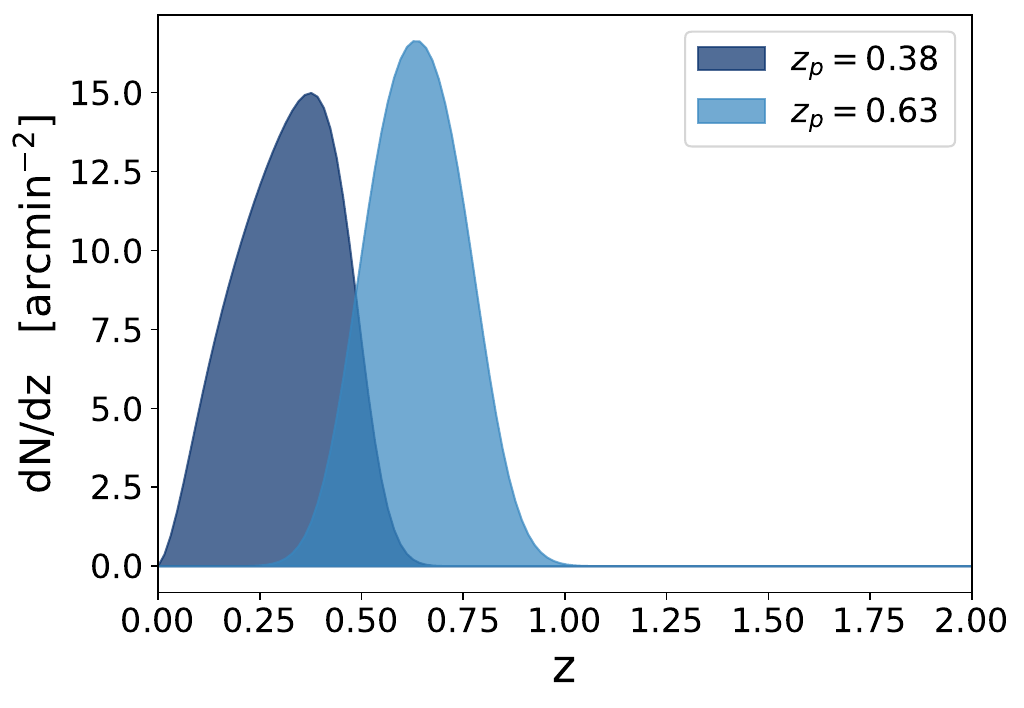}
\hfill
\caption{\label{fig:abacus_bins} Source and maglim lens samples from the DESC Bias Challenge \cite{LSSTDarkEnergyScience:2023qfp} selected for the analyses in this work.}
\end{figure}

The redshift distributions are based on the LSST Y10 sample as defined in the LSST DESC Science Requirements document \cite{LSSTDarkEnergyScience:2023qfp}. The source sample is divided in five redshift bins of equal numbers of galaxies, assuming a photo-z scatter of $\sigma_z = 0.05(1+z)$. The maglim lens sample is generated under the assumption that the same set of galaxies is used for lensing and clustering measurements, which makes the lens and source bins identical. Of the five available bins, we select those with maxima at 0.38 and 0.63 for our estimations (figure \ref{fig:abacus_bins}). As in the previous case (subsection \ref{sec:lik_analysis_sdv}), we select $C^{00}_{gg}$, $C^{00}_{g\kappa}$, $C^{11}_{gg}$, and $C^{11}_{g\kappa}$ as data vectors. For $\ell$-binning, we use the edges: 

\begin{equation*}
\begin{aligned}
    \ell_\mr{edge} \in &\{ 32, 42, 52, 66, 83, 105, 132, 167, 210, 265, 333, 420, 528, \\
    &665, 838, 1054, 1328, 1672, 2104, 2650, 3336, 4200, 5287 \}    .
\end{aligned}
\end{equation*}

The Abacus cosmology flat $\Lambda$CDM specified by the parameters: $\Omega_c = 0.2644$, $\Omega_b = 0.0493$, $\sigma_8 = 0.809$, $h = 0.6736$, and $n_s = 0.9649$. As with the previous case, we treat only $\Omega_c$ and $\sigma_8$ as free cosmological parameters, whereas $b_1$ and shot noise are nuisance parameters. We adopt wide flat priors except on shot noise, which we sample in the range of $\pm30\%$ its Poissonian value. These priors are presented in table \ref{tab:priors_abac}.

\input{priors_abac}

We test whether TE is $\kmax$-independent by performing estimations with $\kmax = 1 \\\hmpc$. However, since in this case the data vectors are unique, we can't evaluate robustness in this respect. As in the previous case, we compare results of TE and scale cuts by matching the tensions and assessing the FoMs. We also adopt a tension threshold of 0.3$\sigma$ for an analysis to be considered successful. To evaluate robustness with respect to changes in the fiducial cosmology, we generate a new envelope and mean error in a different cosmology and apply them to an analysis of the same data vectors. This procedure is different from the one adopted for the other data set (subsection \ref{sec:lik_analysis_sdv}) because, in that case, we could keep the envelope and mean error constant and vary the test vectors instead, which is computationally less expensive.

To evaluate the posteriors in this test and in the case of simulated data vectors, we use the Monte Carlo Markov Chain sampler ``mcmc'' of Cobaya, which is extended from CosmoMC \cite{Lewis_2002, Lewis_2013, Torrado_2021}. For each estimation, we run two chains in four cores each. The level of convergence is given in terms of the Gelman-Rubin statistic as $R-1=0.05$, which corresponds to chains with length between 3,360 and 9,700 steps.

%% file: priors_sdv.tex
\begin{table}
\centering
\label{tab:sdv_table}
\resizebox{0.4\linewidth}{!}{%
\begin{tabular}{|c|c|c|} 
\hhline{|===|}
Parameter                  & Fiducial     & Prior               \\ 
\hline
$\Omega_c$              & 0.266        & Flat(0.20, 0.34)    \\
$\Omega_b$              & 0.05         & Fixed               \\
$\sigma_8$              & 0.831        & Flat(0.60, 0.95)    \\
$h_0$                   & 0.67         & Fixed               \\
$n_s$                   & 0.9645       & Fixed               \\
$b_1$ ($z_p = 0.36$) & 0.7          & Flat(-0.3, 1.7)     \\
$b_1$ ($z_p = 0.84$)   & 1.0          & Flat(0, 2.0)        \\
Shot noise                 & $1/\Bar n$ & Flat($\pm 30\%)$  \\
\hhline{|===|}
\end{tabular}
}\caption{\label{tab:priors_sdv} Priors and fiducial values for the analysis with simulated data vectors.}
\end{table}

%% file: priors_abac.tex
\begin{table}
\centering
\label{tab:sdv_table}
\resizebox{0.4\linewidth}{!}{%
\begin{tabular}{|c|c|c|} 
\hhline{|===|}
Parameter                  & Fiducial     & Prior               \\ 
\hline
$\Omega_c$              & 0.2644       & Flat(0.18, 0.36)    \\
$\Omega_b$              & 0.0493       & Fixed               \\
$\sigma_8$              & 0.809        & Flat(0.60, 0.95)    \\
$h_0$                   & 0.6736       & Fixed               \\
$n_s$                   & 0.9649       & Fixed               \\
$b_1$ ($z_p = 0.38$)    & 0.195        & Flat(-0.53, 0.70)     \\
$b_1$ ($z_p = 0.63$)    & 0.407        & Flat(-0.23, 0.90)        \\
Shot noise                 & $1/\Bar n$ & Flat($\pm 30\%)$  \\
\hhline{|===|}
\end{tabular}
}\caption{\label{tab:priors_abac} Priors and fiducial values for the analysis with data vectors studied in the DESC Bias Challenge.}
\end{table}

%% file: results.tex
\section{Results}
\label{sec:results}

\subsection{Simulated data vectors}
\label{sec:results_sdv}
The results with simulated data vectors are summarized in table \ref{tab:sdv_table}. We use the mock data vectors specified in table \ref{tab:fid_bias}, where ``typical bias'' is the main vector and ``high bias'' and ``low bias'' are chosen to test robustness. 

\begin{figure}[tbp]
\centering % \begin{center}/\end{center} takes some additional vertical space
\includegraphics[width=0.6\textwidth,clip]{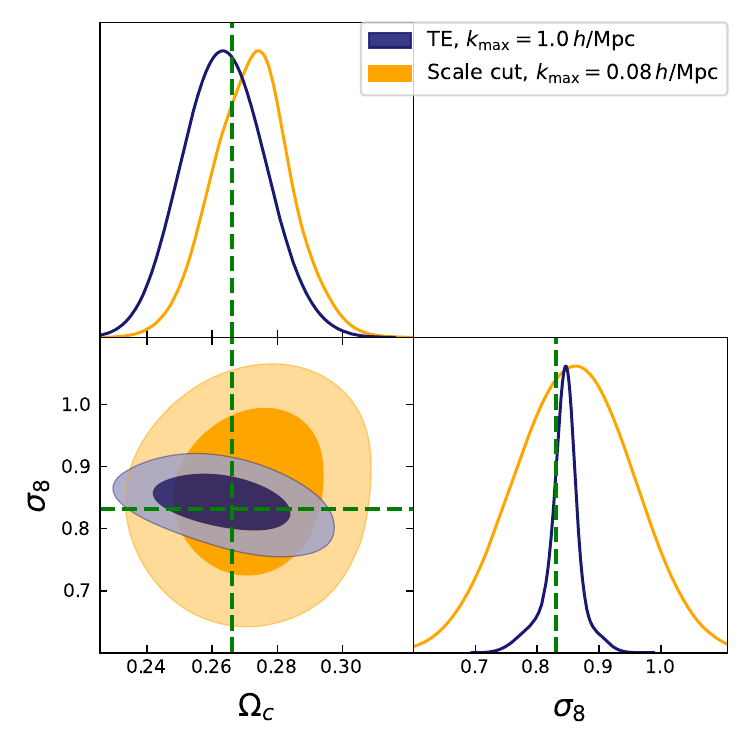}
\hfill
\caption{\label{fig:sdv_result2} Contour plot corresponding to a TE configuration with the ``95\%'' envelope, $\Delta k = 0.05 \hmpc$, and zero $gg$-$g\kappa$ covariance, for the analysis with the ``typical bias'' data vector. The best-corresponding scale cut has $\kmax = 0.08 \hmpc$ (row 1 of table \ref{tab:sdv_table}). TE achieves higher precision for similar tension with fiducial values.}
\end{figure}

\begin{figure}
\centering
\begin{subfigure}{0.50\textwidth}
  \centering
  \includegraphics[width=\linewidth]{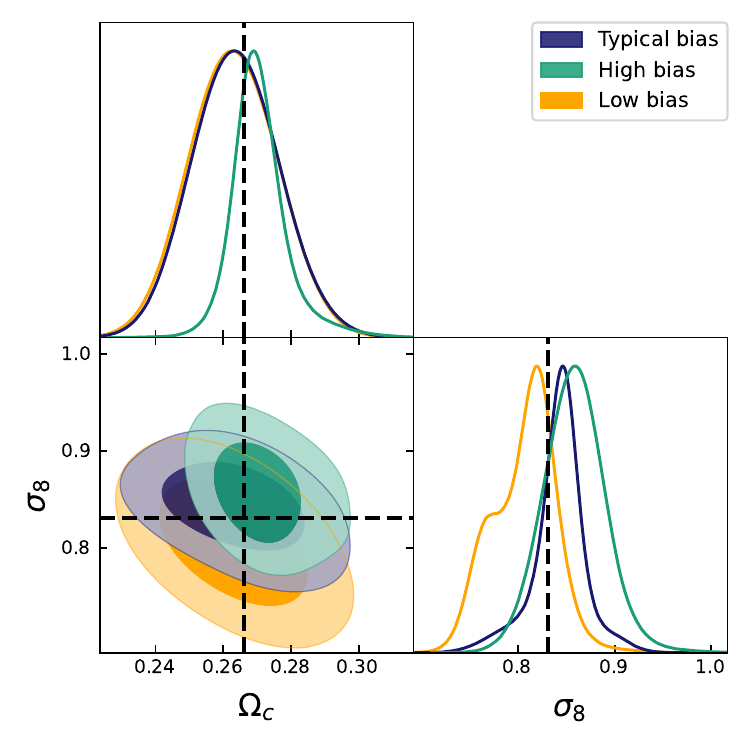}
  \caption{\label{fig:sdv_robust1}``95\%''}
\end{subfigure}%
\begin{subfigure}{0.50\textwidth}
  \centering
  \includegraphics[width=\linewidth]{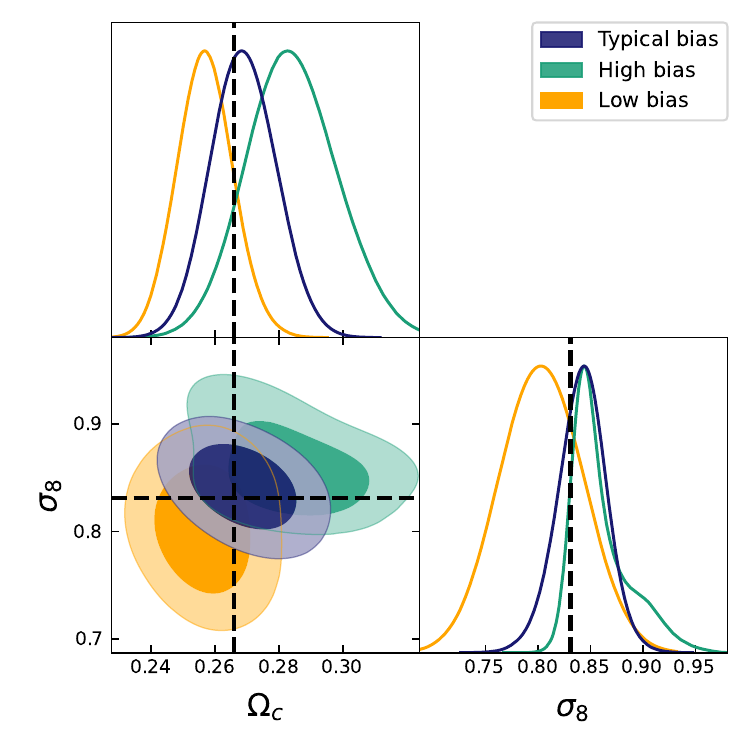}
  \caption{\label{fig:sdv_robust2}``1 sigma''}
\end{subfigure}
\caption{\label{fig:sdv_robust} Contour plots corresponding to TE configurations with the ``95\%'' and ``1 sigma'' envelopes, $\Delta k = 0.05 \hmpc$, and zero $gg$-$g\kappa$ covariance. We show the results of analyses with all three data vectors (rows 1 to 3 of table \ref{tab:sdv_table}, for panel \emph{a}, and rows 4 to 6, for panel \emph{b}). The TE configuration with the ``95\%'' envelope is considered robust under variations of the data vector, as the tension with fiducial values remains within $\simeq 0.7 \sigma$. The analysis with the ``1 sigma'' envelope fails the same test, as here the tension exceeds $1\sigma$ under the ``high bias'' data vector.}
\end{figure}

As apparent from row 1 of the table, the ``95\%'' envelope with $\Delta k = 0.05 \hmpc$ performs better than the scale cut when applied to the ``typical bias'' test vector. It attains higher precision (an FoM of 762 vs. one of 353) for the same level of accuracy, near $0.14\sigma$. Moreover, the tension remains below $\simeq0.7\sigma$ when the data vector is varied to either ``high bias'' or ``low bias'', which shows this configuration passes the test of robustness. We present the corresponding contour plots in figures \ref{fig:sdv_result2} and \ref{fig:sdv_robust1}. The relative results of TE and scale cut with the ``typical bias'' data vector remain qualitatively the same under variations of the fiducial cosmology used to generate the vector, as shown in table \ref{tab:sdv_new_cosmos}.

Since the analyses are made with $k_\mr{max} = 1 \hmpc$ and the results are relatively unbiased, they are considered $\kmax$-independent by the criterion laid out in section \ref{sec:lik_analysis}. This means that high $\ell$-modes are correctly suppressed by the envelopes.

The configuration with the ``1 sigma'' envelope and $\Delta k = 0.05 \hmpc$ also achieves higher precision than the scale cut in the analysis with the ``typical bias'' data vector. However, it exhibits lack of robustness under the change to the ``high bias'' vector, as in this case the tension rises above $1.2\sigma$ (figure \ref{fig:sdv_robust2}).

\input{sdv_table1}

\input{sdv_cosmos}

\subsection{Data vectors used in the DESC Bias Challenge}

Results with data vectors used in the DESC Bias Challenge are summarized in table \ref{tab:abac_table}. With the ``1 sigma'' envelope and $\Delta k$ equal to $0.10$ or $0.30 \hmpc$, TE recovers fiducial parameters with precision considerably higher than the scale cut's for the same level of accuracy. For $\Delta k = 0.05 \hmpc$, the two methods produce comparable results. 

\begin{figure}[tbp]
\centering % \begin{center}/\end{center} takes some additional vertical space
\includegraphics[width=0.6\textwidth,clip]{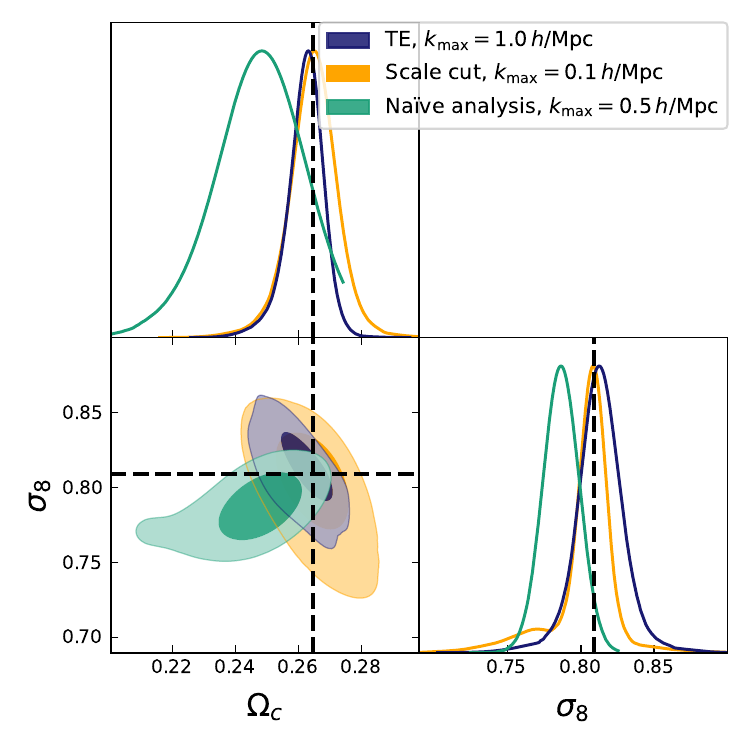}
\hfill
\caption{\label{fig:abac_result1} Contour plot of the $1\sigma$ level corresponding to a TE configuration with the ``1 sigma'' envelope, $\Delta k = 0.10 \hmpc$, and zero $gg$-$g\kappa$ covariance, for the analysis with the data vector used in the DESC Bias Challenge. The best-corresponding scale cut has $\kmax = 0.10 \hmpc$ (row 8 of table \ref{tab:abac_table}). TE achieves higher precision for smaller tension with fiducial values. In contrast, a scale cut at $\kmax = 0.5\hmpc$ produces a tension higher than $1\sigma$ and a reduced $\chi^2$ ($\chi^2$ per degree of freedom) larger than 8. Its poor goodness of fit underlies the spurious inversion of the degeneracy between $\Omega_c$ and $\sigma_8$.}
\end{figure}

Since these analyses are made with $\kmax = 1\hmpc$, we consider them $\kmax$-independ\-ent by the criterion established in section $\ref{sec:lik_analysis}$. An example of triangle plot with TE and scale cut results is shown in figure \ref{fig:abac_result1}. We also include the output of a naïve analysis that assumes linear bias and $\kmax = 0.5 \hmpc$. We find the latter to exhibit large tension with fiducial values and poor goodness-of-fit with the data. In contrast, the addition of TE covariance permits the use of high $\kmax$ without loss of accuracy, as the statistical weight of large $\ell$-modes is smoothly reduced. 

As for robustness under variations of the cosmology, we find that TE produces the same accuracy as in the case where the TE covariance is computed in the data vector's cosmology, but with lower constraining power (table \ref{tab:abac_new_cosmos}). Contrary to the situation observed with simulated data vectors, therefore, with this data set TE doesn't pass this test of robustness, at least with respect to precision.

With the ``95\%'' envelope, TE is still able to recover fiducial parameters with tension lower than 0.2$\sigma$. However, for $\Delta k$ equal to $0.05$ or $0.10 \hmpc$, the precision is markedly lower than the scale cut's, and we consider these configurations unsuccessful. Moreover, if $\Delta k = 0$ or the TE $gg$-$g\kappa$ covariance is nonzero (i.e., computed with formula \eqref{eq:cov_e_specific}), we do not obtain a good fit with TE in the linear-bias model. Indeed, although not shown in the table, we found the inclusion of the TE $gg$-$g\kappa$ covariance to worsen the results of all TE configurations tested with this data set. The inclusion of this component causes both the precision and the estimation error to rise excessively, which points to an inadequacy in the modeling of that matrix. 

This behavior is similar to the dependence of precision on $\Delta k$ as described in section \ref{sec:envelopes}. There, we argued that a large coherence length can cause a spurious rise in precision because of the exclusion of theoretical vectors with sharp variations across $\Delta k$. Similarly in this case, modeling a likelihood function with strong correlation between theoretical errors on $C^{gg}_\ell$ and $C^{g\kappa}_\ell$ can exclude models in which these errors are relatively uncorrelated. Here as well, this represents a reduction of the region in parameter space where the likelihood is dense, or a gain in precision. Moreover, if the supposed large correlation is spurious (unphysical), estimation error increases as well, as models close to the fiducial cosmology (relatively uncorrelated) are excluded. Although uncommon in practice, an increase in precision caused by large off-diagonal covariance elements is theoretically well-founded, as illustrated by means of a toy model in appendix \ref{sec:appendix}. 

To investigate why the model of $\mr{Cov}^\mr{TE}_{gg,g\kappa}(\ell,\ell')$ has proved inadequate, we may recall that it is based on a logical extrapolation from the functional forms of $\mr{Cov}^\mr{TE}_{gg,gg}(\ell,\ell')$ and $\mr{Cov}^\mr{TE}_{g\kappa,g\kappa}(\ell,\ell')$. That is:

\begin{equation}
    \mr{Cov}^\mr{TE}_{gg,g\kappa}(\ell,\ell') = E^{gg}_\ell E^{g\kappa}_{\ell'} \rho_{\ell\ell'} ,
\end{equation}

\noindent where $E^{gg}_\ell$ and $E^{g\kappa}_\ell$ are the envelopes corresponding to the spectra $C^{gg}_\ell$ and $C^{g\kappa}_\ell$ and $\rho_{\ell\ell'}$ are the correlation coefficients. We construct the envelopes from the standard deviation (or approximately twice its value, in the case of the ``95\%'' envelope) of the theoretical error. As noted in section \ref{sec:envelopes}, for $\ell = \ell'$ the equation above expresses the assumption: $\sigma^2_{XY} \simeq \sigma_X \sigma_Y$, where $X$ and $Y$ are random variables with covariance $\sigma^2_{XY}$ and standard deviations $\sigma_X$ and $\sigma_Y$. This assumption is, at best, a good approximation. Indeed, several earlier works \cite{Chudaykin_2021, baldauf2016lss, Moreira_2021} have neglected to model the covariance between theoretical errors on different spectra. When some assumption in this respect was necessary (e.g., on the relation between spectra at different redshift bins), the covariance was set to zero. 

By setting the $gg$-$g\kappa$ TE covariance to zero in this work as well, we can recover the fiducial parameters within $0.2\sigma$ tension, as shown in table \ref{tab:abac_table}. This choice also causes a significant loss of precision, however, likely because completely uncorrelated errors on $C^{gg}_\ell$ and $C^{g\kappa}_\ell$ are now permitted by the model. This may explain the relatively poor performance of TE with ``95\%'' envelope. Setting the covariance to zero also affects the results with the ``1 sigma'' envelope, but in this case precision is sufficiently high (due to the envelope's smaller size) for them to surpass the scale cut's output. It seems clear that a more judicious modeling of this part of the covariance becomes necessary in 2$\times$2- and 3$\times$2-point analyses.

The choice of $\Delta k$ is significant as well. Both the value that minimizes the SNR ($\Delta k = 0.05 \hmpc$) and the value that ensures correlation at the BAO scale ($\Delta k = 0.1 \hmpc$) \cite{Chudaykin_2021} produce comparatively good results. However, setting $\Delta k$ to zero -- i.e., treating the errors at each $\ell$ mode as independent, which is also prone to increase precision -- causes a considerable loss of accuracy. Indeed, we don't achieve a good fit with the linear-bias model in this case either.

In summary, a TE configuration with the ``1 sigma'' envelope and $\Delta k$ equal to $0.05$ or $0.10 \hmpc$ produces good results in precision and accuracy when compared to the scale cut. We connect the relatively poor results under the ``95\%'' envelope to an inadequate modeling of the $gg$-$g\kappa$ TE covariance. Setting it to zero allows the recovery of unbiased cosmology with the linear-bias model, although the resulting precision loss makes the ``95\%'' envelope less viable than a scale cut.

\input{abac_table1}

\input{abac_cosmos}

%% file: sdv_table1.tex
\begin{table}
\centering
\label{tab:sdv_table}
\resizebox{\linewidth}{!}{%
\begin{tabular}{|c|c|c|c|c|c|c|c|c|c|c|c|c|c|} 
\hhline{|==============|}
\multirow{2}{*}{\begin{tabular}[c]{@{}c@{}}Row\\number\end{tabular}} & \multicolumn{7}{c|}{Theoretical error}                                                                                                                                                     & \multicolumn{5}{c|}{Best corresponding scale cut}                                                          & \multirow{2}{*}{\begin{tabular}[c]{@{}c@{}}Preferable\\ method\end{tabular}}  \\ 
\cline{2-13}
                                                                     & \begin{tabular}[c]{@{}c@{}}TE gg-g$\kappa$\\ covariance\end{tabular} & \begin{tabular}[c]{@{}c@{}}$\Delta k$ \\ {[}$h/$Mpc]\end{tabular} & Envelope & Test vector  & FoM  & Tension & FoB  & \begin{tabular}[c]{@{}c@{}}$k_\mathrm{max}$\\ {[}$h/$Mpc]\end{tabular} & Test vector  & FoM & Tension & FoB  &                                                                               \\ 
\hline
1                                                                    & Zero                                                                 & 0.05                                                              & 95\%     & Typical bias & 762  & 0.14    & 0.41 & 0.08                                                                 & Typical bias & 353 & 0.18    & 0.44 & TE                                                                            \\
2                                                                    & Zero                                                                 & 0.05                                                              & 95\%     & High bias    & 940  & 0.64    & 1.27 & 0.08                                                                 & High bias    & 277 & 0.05    & 0.52 & TE                                                                            \\
3                                                                    & Zero                                                                 & 0.05                                                              & 95\%     & Low bias     & 687  & 0.32    & 1.04 & 0.08                                                                 & Low bias     & 518 & 0.23    & 0.32 & TE                                                                            \\
4                                                                    & Zero                                                                 & 0.05                                                              & 1 sigma  & Typical bias & 1186 & 0.20    & 0.73 & 0.08                                                                 & Typical bias & 353 & 0.18    & 0.44 & Scale cut                                                                     \\
5                                                                    & Zero                                                                 & 0.05                                                              & 1 sigma  & High bias    & 651  & 1.28    & 2.02 & 0.08                                                                 & High bias    & 277 & 0.05    & 0.52 & Scale cut                                                                     \\
6                                                                    & Zero                                                                 & 0.05                                                              & 1 sigma  & Low bias     & 803  & 0.73    & 1.46 & 0.08                                                                 & Low bias     & 518 & 0.23    & 0.32 & Scale cut                                                                     \\
7                                                                    & Zero                                                                 & 0.10                                                              & 95\%     & Typical bias & 829  & 0.11    & 0.38 & 0.09                                                                 & Typical bias & 512 & 0.08    & 0.56 & TE                                                                            \\
8                                                                    & Zero                                                                 & 0.30                                                              & 95\%     & Typical bias & 896  & 0.19    & 0.58 & 0.09                                                                 & Typical bias & 512 & 0.08    & 0.56 & TE                                                                            \\
9                                                                    & Nonzero                                                              & 0.05                                                              & 95\%     & Typical bias & 2278 & 0.12    & 0.38 & 0.09                                                                 & Typical bias & 512 & 0.08    & 0.56 & TE                                                                            \\
10                                                                   & Zero                                                                 & 0.10                                                              & 1 sigma  & Typical bias & 1072 & 0.19    & 0.70 & 0.09                                                                 & Typical bias & 512 & 0.08    & 0.56 & TE                                                                            \\
11                                                                   & Zero                                                                 & 0.30                                                              & 1 sigma  & Typical bias & 1275 & 0.24    & 0.70 & 0.09                                                                 & Typical bias & 512 & 0.08    & 0.56 & TE                                                                            \\
\hhline{|==============|}
\end{tabular}
}\caption{\label{tab:sdv_table} Comparison of estimation results with theoretical error and the usual method of scale cuts, for simulated data vectors. For estimations with TE, we take $\kmax = 1 \hmpc$. We use ``typical bias'' as main data vector and ``high bias'' and ``low bias'' to test robustness. As apparent from rows 1 to 3, a TE configuration with the ``95\%'' envelope, $\Delta k = 0.05 \hmpc$, and zero gg-g$\kappa$ covariance exhibits higher precision than the scale cut for the same level of accuracy, under the ``typical bias'' data vector. It also passes the test of robustness, as the tension remains within $\simeq 0.7\sigma$ when the data vector is varied. A configuration with the ``1 sigma'' envelope (rows 4 to 6) fails the latter test.}
\end{table}

%% file: sdv_cosmos.tex
\begin{table}[]
\centering
\resizebox{0.6\columnwidth}{!}{%
\begin{tabular}{|cc|ccc|cccc|}
\hhline{|=========|}
\multicolumn{2}{|c|}{\begin{tabular}[c]{@{}c@{}}Cosmological \\ parameter\end{tabular}} & \multicolumn{3}{c|}{Theoretical error}                          & \multicolumn{4}{c|}{Scale cut}                                                                                                                                   \\ \hline
\multicolumn{1}{|c|}{$\Omega_c$}                         & $\sigma_8$                        & \multicolumn{1}{c|}{FoM}  & \multicolumn{1}{c|}{Tension} & FoB  & \multicolumn{1}{c|}{\begin{tabular}[c]{@{}c@{}}$k_\mathrm{max}$\\ {[}$h/$Mpc{]}\end{tabular}} & \multicolumn{1}{c|}{FoM}   & \multicolumn{1}{c|}{Tension} & FoB  \\ \hline
\multicolumn{1}{|c|}{0.266}                            & 0.831                           & \multicolumn{1}{c|}{762}  & \multicolumn{1}{c|}{0.14}    & 0.41 & \multicolumn{1}{c|}{0.08}                                                                     & \multicolumn{1}{c|}{353} & \multicolumn{1}{c|}{0.18}    & 0.44 \\
\multicolumn{1}{|c|}{0.302}                            & 0.819                           & \multicolumn{1}{c|}{1113} & \multicolumn{1}{c|}{0.48}    & 0.87 & \multicolumn{1}{c|}{0.08}                                                                     & \multicolumn{1}{c|}{655} & \multicolumn{1}{c|}{0.70}     & 0.90 \\
\multicolumn{1}{|c|}{0.230}                            & 0.837                           & \multicolumn{1}{c|}{755}  & \multicolumn{1}{c|}{0.06}    & 0.50 & \multicolumn{1}{c|}{0.08}                                                                     & \multicolumn{1}{c|}{278} & \multicolumn{1}{c|}{0.74}    & 0.66 \\ \hhline{|=========|}
\end{tabular}%
}\caption{\label{tab:sdv_new_cosmos} Estimation results of TE and scale cut for several cosmologies. In all three cases, we use the envelope and mean theoretical error calculated in the fiducial cosmology $\Omega_c = 0.266$ and $\sigma_8 = 0.831$. We apply them to data vectors generated in the same (row 1) and in two different cosmologies (rows 2 and 3). We use the ``typical bias'' parameters for the data vector in all cases. This test corresponds to a TE configuration with the ``95\%'' envelope, $\Delta k = 0.05 \hmpc$, and zero TE gg-g$\kappa$ covariance. TE produces higher precision than the scale cut for similar or lower tension under variations of the cosmological parameters.}
\end{table}

%% file: abac_table1.tex
\begin{table}
\centering
\label{tab:abac_table}
\resizebox{\linewidth}{!}{%
\begin{tabular}{|c|c|c|c|c|c|c|c|c|c|c|c|} 
\hhline{|============|}
\multirow{2}{*}{\begin{tabular}[c]{@{}c@{}}Row\\number\end{tabular}} & \multicolumn{6}{c|}{Theoretical error}                                                                                                                                     & \multicolumn{4}{c|}{Best corresponding scale cut}                                              & \multirow{2}{*}{\begin{tabular}[c]{@{}c@{}}Preferable\\ method\end{tabular}}  \\ 
\cline{2-11}
                                                                     & \begin{tabular}[c]{@{}c@{}}TE gg-g$\kappa$\\ covariance\end{tabular} & \begin{tabular}[c]{@{}c@{}}$\Delta k$\\ {[}$h$/Mpc]\end{tabular} & Envelope & FoM  & Tension & FoB  & \begin{tabular}[c]{@{}c@{}}$k_\mathrm{max}$\\ {[}$h$/Mpc]\end{tabular} & FoM  & Tension & FoB  &                                                                               \\ 
\hline
1                                                                    & Zero                                                                 & 0                                                                & 95\%     & –    & –       & –    & –                                                                      & –    & –       & –    & Scale cut                                                                     \\
2                                                                    & Zero                                                                 & 0.05                                                             & 95\%     & 1481 & 0.13    & 0.60 & 0.10                                                                   & 2292 & 0.14    & 0.67 & Scale cut                                                                     \\
3                                                                    & Zero                                                                 & 0.10                                                             & 95\%     & 1588 & 0.17    & 0.37 & 0.10                                                                   & 2292 & 0.14    & 0.67 & Scale cut                                                                     \\
4                                                                    & Zero                                                                 & 0.30                                                             & 95\%     & 2267 & 0.17    & 0.68 & 0.10                                                                   & 2292 & 0.14    & 0.67 & Scale cut                                                                     \\
5                                                                    & Nonzero                                                              & 0.05                                                             & 95\%     & –    & –       & –    & –                                                                      & –    & –       & –    & Scale cut                                                                     \\
6                                                                    & Zero                                                                 & 0                                                                & 1 sigma  & –    & –       & –    & –                                                                      & –    & –       & –    & Scale cut                                                                     \\
7                                                                    & Zero                                                                 & 0.05                                                             & 1 sigma  & 1708 & 0.04    & 0.45 & 0.08                                                                   & 1722 & 0.01    & 0.29 & Either                                                                        \\
8                                                                    & Zero                                                                 & 0.10                                                             & 1 sigma  & 3602 & 0.06    & 0.44 & 0.10                                                                   & 2292 & 0.14    & 0.67 & TE                                                                            \\
9                                                                    & Zero                                                                 & 0.30                                                             & 1 sigma  & 4784 & 0.30    & 1.03 & 0.10                                                                   & 2292 & 0.14    & 0.67 & TE                                                                            \\
\hhline{|============|}
\end{tabular}
}\caption{\label{tab:abac_table} Comparison of estimation results with TE and scale cuts, for data vectors used in the DESC Bias Challenge. Two TE configurations with the ``1 sigma'' envelope and zero gg-g$\kappa$ covariance achieve higher precision than the scale cut for similar or smaller tension. Configurations with the ``95\%'' envelope are outperformed at least marginally by the scale cuts. We don't obtain a good fit if $\Delta k = 0$ or if the TE gg-g$\kappa$ covariance is nonzero, i.e., computed with formula \eqref{eq:cov_e_specific}. These poor fits with TE correspond to the missing rows on the table, where we note ``scale cut'' as preferable method because any reasonable choice of $\kmax$ produces a better fit.}
\end{table}

%% file: abac_cosmos.tex
\begin{table}[]
\centering
\resizebox{0.65\columnwidth}{!}{%
\begin{tabular}{|cc|ccc|cccc|}
\hhline{|=========|}
\multicolumn{2}{|c|}{\begin{tabular}[c]{@{}c@{}}Envelope's \\ cosmology\end{tabular}} & \multicolumn{3}{c|}{Theoretical error}                          & \multicolumn{4}{c|}{Scale cut}                                                                          \\ \hline
\multicolumn{1}{|c|}{$\Omega_c$}                     & $\sigma_8$                     & \multicolumn{1}{c|}{FoM}  & \multicolumn{1}{c|}{Tension} & FoB  & \multicolumn{1}{c|}{$k_\mathrm{max}$ [$h/$Mpc]} & \multicolumn{1}{c|}{FoM}  & \multicolumn{1}{c|}{Tension} & FoB  \\ \hline
\multicolumn{1}{|c|}{0.264}                          & 0.809                          & \multicolumn{1}{c|}{1481} & \multicolumn{1}{c|}{0.13}    & 0.60 & \multicolumn{1}{c|}{0.10}             & \multicolumn{1}{c|}{2292} & \multicolumn{1}{c|}{0.14}    & 0.67 \\
\multicolumn{1}{|c|}{0.300}                          & 0.797                          & \multicolumn{1}{c|}{571}  & \multicolumn{1}{c|}{0.12}    & 0.75 & \multicolumn{1}{c|}{0.10}             & \multicolumn{1}{c|}{2292}  & \multicolumn{1}{c|}{0.14}    & 0.67 \\ \hhline{|=========|}
\end{tabular}%
}\caption{\label{tab:abac_new_cosmos} Results of TE and scale cuts for two cosmologies. To obtain the results shown in row 1, we compute the envelope and mean error in the Abacus cosmology ($\Omega_c = 0.264$ and $\sigma_8 = 0.809$), i.e., the same cosmology that underlies the DESC Bias Challenge data vector. Row 2 shows the case where the envelope and mean error are computed with a different set of cosmological parameters, obtained by varying $\Omega_c$ and $\sigma_8$ approximately within the 95\% confidence interval of recent observational bounds \cite{Planck2018}. In both cases we use a TE configuration with the ``95\%'' envelope, $\Delta k = 0.05 \hmpc$, and zero gg-g$\kappa$ covariance. TE achieves similar tension in the new cosmology, but with precision significantly lower.}
\end{table}

%% file: conclusion.tex
\section{Conclusion}
\label{sec:conc}

In this work, we investigate a method of statistical analysis of large-scale structure that takes account of theoretical uncertainty. We consider the uncertainty arising from neglecting higher-order operators in the galaxy-bias expansion. Since it smoothly increases as the spatial scales decrease, including this uncertainty in the analysis can in principle replace sharp scale cuts by rendering unnecessary a calibration of $k_\mr{max}$. Moreover, it possibly attains higher constraining power in view of the inclusion of higher $k$-modes, albeit with reduced statistical weight. In comparison with an estimation with the full-bias model, this approach has the advantage of requiring a smaller parameter space, which implies lower computational cost and alleviated effects of volume projection. 

We apply this method to a simplified setting of 2$\times$2-point analyses of galaxy clustering and galaxy-galaxy lensing. In this context, we call ``theoretical error'' (TE) the difference between linear-bias and nonlinear-bias (to second order) predictions for the relevant angular spectra. Since theoretical uncertainty lacks an unambiguous definition, we make several modeling choices when specifying the TE likelihood. Our model is thus determined by an ``envelope'', a correlation length, and a mean theoretical error. We define the envelope as the square root of the diagonal of the TE covariance matrix, hence the standard deviation of TE at a given $\ell$ mode. We construct it by sampling higher-order bias parameters from simulation-based distributions. These same distributions also determine the mean theoretical error. The correlation length requires an independent ansatz, and we test several possibilities: setting it to zero, choosing a value that ensures correlation at the typical BAO scale ($k_\mr{BAO}$), or selecting a conservative value that eliminates spurious precision. One of our findings is that the correlation length sensitively influences the results of TE. It must be chosen judiciously when the data vectors lack a clear characteristic scale such as $\Delta k_{\rm BAO}$, which arises naturally in the context of spectroscopic surveys.

To assess the viability of the TE approach, we apply it to mock likelihood analyses of 2$\times$2-point spectra. We use two sets of mock data vectors: (i) simulated data vectors constructed from the same distributions used to generate the TE envelope, and (ii) data vectors previously studied in the LSST DESC Bias Challenge \cite{LSSTDarkEnergyScience:2023qfp}, which are derived from {\tt AbacusSummit}, a high-fidelity $N$-body simulation suite \cite{Garrison_2018}.

We compare TE and scale cut in terms of precision, accuracy, and (for the first data set) robustness under variations of the data vector. In most cases we show that the two methods yield comparable results, the relative performance depending mostly on the choice of the TE envelope. With simulated data vectors, we find TE configurations that attain precision significantly higher than the scale cuts with the same accuracy level. This performance is robust under variations of the data vectors and cosmological parameters. With vectors from the DESC Bias Challenge, we also obtain a number of TE configurations that exhibit higher constraining power than the scale cut for similar accuracy. However, the scale cut yields higher precision in most cases, and TE fails a test of robustness under variations of cosmological parameters. 

We relate the comparatively poor results of TE with the second dataset to an unsatisfactory modeling of the $gg$-$g\kappa$ TE covariance, which involves difficulties not encountered in the $gg$-$gg$ and $g\kappa$-$g\kappa$ cases. Namely, it is not clear how to rigorously extrapolate it from the functional form adopted for the other matrices. Indeed, past works dedicated to studying theoretical error covariances haven't dealt with the cross-probe contribution \cite{baldauf2016lss, Moreira_2021, Chudaykin_2021}. We find that setting it to zero produces accuracy and precision comparable to the scale cut's in some cases, which suggests that a more sophisticated model for this component could considerably increase the performance of TE in cross-probe estimations. This is also true of the covariance between theoretical errors at different redshift bins, which we have neglected in this work in view of the large bin separation. The inclusion of this part of the covariance, through some generalization of eq. \eqref{eq:cov_e}, is recommended for further work.

In view of these results, we consider TE a promising technique in tomographic 2$\times$2- and 3$\times$2-point analyses. We emphasize the advantage of its not requiring a choice of $\kmax$, as the augmented covariance matrix suppresses small-scale modes. With respect to estimations with nonlinear bias models, the smaller parameter space entails lower computational cost and less severe projection effects. The method's insufficient performance in some of the tests, however, illustrates the need for more judicious modeling choices, especially concerning the theoretical error cross-probe covariance. 

Other potential extensions to our work include: the generalization to the case of cross-correlations with CMB lensing, Compton-$y$ maps from the thermal Sunyaev-Zel'dovich effect~\cite{Gatti_2022, Pandey_2022}, and other probes commonly cross-correlated with galaxies; an analysis with data vectors from the DESC Bias Challenge using the full redshift-bin structure of the LSST, including 10 bins of lens samples and 5 bins of source samples; the inclusion of theoretical-error covariance from intrinsic alignment modeling in the case of cosmic shear~\cite{chen2024lagrangiantheorygalaxyshape}; and, potentially, the treatment of theoretical error from higher-than-second-order biases, as current analyses have now commonly begun to employ the second-order Lagrangian bias model as a fiducial baseline~\cite{White_2022, Chen_2022, sailer2024cosmologicalconstraintscrosscorrelationdesi, chen2024lensinglowanalysisdesitimesdes}.

%% file: noise_table.tex
\begin{table}
\centering
\refstepcounter{table}
\label{tab:abac_table}
\resizebox{0.75\linewidth}{!}{%
\begin{tabular}{|c|c|c|c|c|c|c|c|c|} 
\hhline{|=========|}
\multirow{2}{*}{\begin{tabular}[c]{@{}c@{}}Data\\vector\end{tabular}} & \multicolumn{4}{c|}{TE}                                                               & \multicolumn{4}{c|}{Scale cut}                                                                                                         \\ 
\cline{2-9}
                                                                      & Tension & FoM   & FoB  & \begin{tabular}[c]{@{}c@{}}Reduced $\chi^2$ \end{tabular} & Tension               & FoM                    & FoB                   & \begin{tabular}[c]{@{}c@{}}Reduced $\chi^2$\end{tabular}  \\ 
\hline
No noise                                                              & 0.14    & 761.6 & 0.41 & 0.00                                                         & 0.18                  & 353.5                  & 0.44                  & 0.00                                                          \\
Noise 1                                                               & 0.67    & 594.3 & 1.24 & 0.66                                                         & 0.51                  & 208.9                  & 1.44                  & 1.70                                                          \\
Noise 2                                                               & 0.70    & 492.5 & 1.33 & 0.37                                                         & 0.80                  & 169.7                  & 1.58                  & 0.63                                                          \\
Noise 3                                                               & 0.06    & 468.3 & 0.11 & 0.56                                                         & 0.50                  & 278.8                  & 0.79                  & 1.48                                                          \\
Noise 4                                                               & 0.21    & 476.2 & 0.75 & 0.35                                                         & 0.55                  & 257.0                  & 1.10                  & 0.58                                                          \\
Noise 5                                                               & 0.48    & 592.1 & 1.05 & 0.69                                                         & 0.46                  & 234.14                 & 0.96       & 2.04                                         \\
\hhline{|=========|}
\end{tabular}
}\caption{\label{tab:noise} Results of TE and scale cut for noisy realizations of the ``typical bias'' data vector. The TE configuration is specified by the ``95\%'' envelope, $\Delta k = 0.05 \hmpc$, 
zero $gg$--$g\kappa$ TE covariance, and $\kmax = 1\hmpc$. The scale cut has $\kmax = 0.08\hmpc$. TE exhibits higher robustness to noise. Its reduced $\chi^2$ statistic remains below or equal to 0.7 for all tested data vectors, whereas for the
scale cut it exceeds 1.4 in three of the five cases. TE attains greater precision for similar or higher accuracy.}
\end{table}